\documentstyle[eqsecnum,multicol,epsfig,aps,prl,array]{revtex}
\newcommand{\bleq}{\ifpreprintsty
                   \else
                   \end{multicols}\widetext \vspace*{-3.5ex}{\tiny
                   
                \noindent\begin{tabular}[t]{c|}
                   \parbox{0.493\hsize}{~} \\ \hline \end{tabular}}
                                      \fi}
\newcommand{\eleq}{\ifpreprintsty
                   \else
                   {\tiny\hspace*{\fill}\begin{tabular}[t]{|c}\hline
                    \parbox{0.49\hsize}{~} \\
                    \end{tabular}}\vspace*{-2.5ex}\begin{multicols}{2}
                    \narrowtext
                    \fi}
\newcommand{\bcols}{\ifpreprintsty\else\begin{multicols}{2} 
        \narrowtext\fi}
\newcommand{\ecols}{\ifpreprintsty\else\end{multicols}\fi}

\widetext
\begin{document}
\title{Charged domain walls as quantum strings living on a lattice.}
\author{Henk Eskes$^*$, Osman Yousif Osman, Rob Grimberg,
        Wim van Saarloos and Jan Zaanen}
\address{Instituut-Lorentz, Leiden University, P.O. Box 9506,
         NL-2300 RA Leiden, The Netherlands} 

\date{\today} \maketitle
\begin{abstract}
  Recently experimental evidence is accumulating that the the doped  
  holes in the high-T$_c$ cuprate superconductors form domain walls
  separating antiferromagnetic domains. These so-called stripes are
  line-like objects and if these persist in the superconducting state,
  high-$T_c$ superconductivity is related to a quantum string liquid.
  In this paper the problem of a single quantum meandering string
  living on a lattice is considered.  A kink model is introduced for
  the string dynamics which allows us to analyze lattice
  commensuration aspects. Building on earlier work by den Nijs and
  Rommelse, this lattice string model can be related both to
  restricted Solid-on-Solid models, describing the worldsheet of the
  string in Euclidean space time, and to one-dimensional quantum spin
  chains. At 
  zero-temperature a strong tendency towards orientational order is
  found, and the remaining directed string problem can be treated in
  detail. Quantum delocalized strings are found whose long wavelength
  wandering fluctuation are described by free field theory, and it is
  argued that the fact that the critical phase of delocalized lattice
  strings corresponds to a free Gaussian theory is a very general
  consequence of the presence of a lattice. In addition, the mapping
  on the surface problem is exploited to show the existence of new
  types of localized string phases; some of these are characterized by
  a proliferation of kinks, but the kink flavors are condensed so that
  the long wavelength fluctuations of these strings are suppressed.
  The simplest phase of this kind is equivalent to the incompressible
  (Haldane) phase of the $S=1$ spin chain, and corresponds with a bond
  centered string: the average string position is centered on bonds.
  We also find localized phases of this type which take arbitrary
  orientations relative to the underlying lattice. The possible
  relevance of these lattice strings for the stripes in
  cuprates is discussed. 
\end{abstract}


\bcols

% From bellow is the original Henk heading
%%\documentstyle[prb,aps,multicol,preprint]{revtex}
%\documentstyle[prb,aps,multicol,epsf]{revtex}

%\begin{document}
%\draft
%\title{Charged domain walls as quantum strings living on a lattice.}
%\author{Henk Eskes$^*$, Osman Yousif Osman, Rob Grimberg,
%        Wim van Saarloos and Jan Zaanen}
%\address{Instituut-Lorentz, Leiden University, P.O. Box 9506,
%         NL-2300 RA Leiden, The Netherlands} 
%\date{\today}
%\maketitle

%\begin{abstract} 
%\end{abstract}

%\pacs{}

%\begin{multicols}{2}

\section{Introduction}\label{sec_intro}

A series of experimental developments has changed the perspective on
the problem of high $T_c$ superconductivity drastically. As long as
the doping level is not too high, electrons bind at temperatures well
above $T_c$\cite{pes}, and the superconducting state appears to be in
tight competition with some collective insulating
state\cite{boebinger}. There exists compelling evidence that this
insulating state corresponds with a novel type of electron crystal,
characterized by both spin- and charge condensation: the stripe
phase\cite{tranquada,tranqua1,yamada}. This phase consists of a
regular array of {\em charged magnetic domain walls}: the holes
introduced by doping form line-like textures which are at the same
time anti-phase boundaries, separating antiferromagnetic spin domains
--- see Fig.\ 1{\em a}.
 This stripe phase is observed in systems
where the insulating state is stabilized by $Zn$ doping \cite{yamada,yamada1}
or by the so-called LTT collective pinning
potential\cite{tranquada,tranqua1}.

Inelastic neutron scattering data reveal that strong dynamical stripe
correlations persist in the metallic- and superconducting
regimes\cite{tranqua1,yamada1,tranquada2,mook}.  Although no static stripe
order is present, the magnetic fluctuations as measured by inelastic
neutron scattering should reflect stripe correlations. As was shown
very recently, the magnetic modulation wavevector of the static
stripe phase seems identical with that of the dynamical spin
fluctuations in the metal- and superconductor for various doping
levels\cite{tranqua1}. In addition, it was argued that the anomalous
normal state magnetic dynamics can be explained in terms of domain
wall meandering dynamics\cite{zhvs}.

The exciting possibility arises that the zero temperature
superconducting state is at the same time a relatively mildly
fluctuating quantum stripe fluid. Unlike the rather featureless
diagonal sector of, e.g., $^4He$\cite{ceperley}, it can be imagined
that the charge- and spin sectors of this quantum stripe problem have
an interesting internal structure. Because charged domain walls are
line-like objects, the charge sector might be looked at as a {\em
  quantum string liquid}\cite{zhvs,eskes1,nawi}. Little is known in
general about such problems, and theoretical analysis is needed.  In
order to address the problem of many interacting strings, it is first
necessary to find out the physics of a single string or charged domain
wall in isolation. A string is an extended object, carrying a
non-trivial collective dynamics --- in contrast to particle-like
problems, the elementary constituent of the string liquid poses
already a serious problem. The physics of quantum strings is a rich
subject. This is most easily discussed in terms of path-integrals. In
D+1 dimensional Euclidean space time, a particle corresponds with a
world line, and so the quantum string corresponds with a
``worldsheet''. The statistical physics of membranes is a rich
subject, which is still under active investigation\cite{stringgen}.
 
The debate on the microscopic origin of the stripe instability is far
from closed\cite{zagu,harfo,emkiv,prelov,zaol,nawi,whisca}.
Nevertheless, in this paper we will attempt to isolate some
characteristics which might be common to all present proposals for the
microscopy, to arrive at some general considerations regarding the
quantum meandering dynamics.  From those we will abstract a minimal
model for the string dynamics.  The phase diagram of this model can be
mapped out completely, and turns out to be remarkably rich.

\begin{figure}[t]
%\epsfxsize=0.5\hsize

%\vspace{0.5ex}
%\epsffile{fig_walls.eps}   
\hspace{-2ex}\epsfig{figure=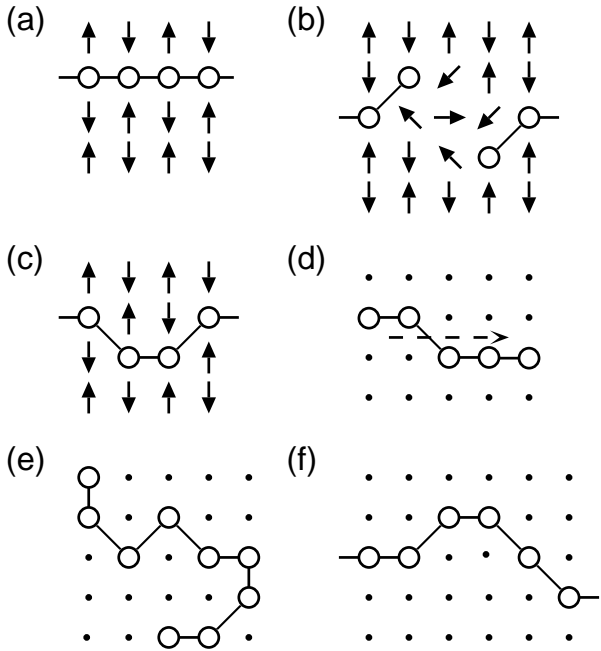,width=8cm}   
%\vspace{1.5ex}
\caption{ (a) A charged domain wall separating spin domains 
of opposite AFM order parameter. (b) Breaking up domain 
walls  causes spin frustration, while (c) ``kinks'' do not. (d)  
 Kinks can gain kinetic energy by moving along the domain   
wall. (e) A typical rough wall. (f) An example of a directed 
 string. }
\label{fig_walls}
\end{figure}

These characteristic features are: {\em (i)} It is assumed that the
charge carriers are confined to domain walls. This is the major
limitation of the present work and it is hoped that at least some
general characteristics of this strong coupling regime survive in the
likely less strongly coupled regime where nature appears to be. {\em
  (ii)} In addition, we assume that domain walls are not broken up, as
sketched in Fig.\ 1{\em b}, as this would lead to strong spin
frustration.  {\em (iii)} Most importantly, we assume a dominant role
of lattice commensuration on the scale of the lattice constant.
Configuration space is built from strings which consist of ``holes''
which live on the sites of an underlying lattice. An example of such a
string configuration is sketched in Fig.\ 1{\em c}. This automatically
implies that the microscopic dynamics is that of {\em kinks} along the
string (Figs.\ 1~{\em c,d}), and this leads to major simplifications
with regard to the long wavelength behavior of the string as a whole.
Note that there is ample evidence for the importance of lattice
commensuration: the scaling of the incommensurability with hole
density $x$ for $x < 1/8$\cite{yamada1}, the special stability at
$x=1/8$\cite{tranqua1}, the LTT pinning mechanism\cite{tranquada}.
{\em (iv)} It is assumed that the strings do not carry other low lying
internal degrees of freedom, apart from the shape fluctuations.
Physically this means that localized strings would be electronic
insulators.  The data of Yamada {\em et al}.\cite{yamada} indicate
that this might well be the case at dopings $x \le 1/8$ (the linear
dependence of the incommensurability on $x$ indicates an on-domain
wall charge commensuration), but it is definitely violated at larger
dopings where the strings should be
metallic\cite{whisca,metstr,Karlsruhe,dhlee}.  Work is in progress on
fluctuating metallic strings, where we find indications that the
collective string dynamics is quite similar to what is presented here
for insulating strings\cite{unpub}.

Given these requirements, one would like to consider a quantum sine
Gordon model\cite{qusine} for the string dynamics,
\begin{equation}
  H = {{1}\over{2}} \int dl \left[ \Pi(l)^2 + { 1 \over {c^2} } \left(
  { {\delta z(l)} \over {\delta l} } \right)^2 + g \sin \left( { {2
      \pi z(l)} \over {a} } \right) \right]~.
\label{qusigo}
\end{equation}
Here $z(l)$ is the transversal displacement at point $l$ on the
string, and $\Pi(l)$ its conjugate momentum defined through the
commution relation $ \left[ \Pi(l), z(l') \right] = i \delta (l -
l')$, and $c$ is the transversal sound velocity. The first two terms
in Eq. (\ref{qusigo}) describe a free string, while the last term is
reponsible for the lattice commensuration effects: every time the
string is displaced by a lattice constant, the potential energy is at
a minimum. This model is well understood\cite{qusine}. When the
strength of the nonlinear interaction exceeds a critical value ($g >
g_c$), the interaction term is relevant and the string localizes. The
excitation spectrum develops a gap and it is characterized by
well-defined kink and anti-kink excitations. When $g < g_c$ the sine
term is irrelevant, and although the dynamics is at least initially
kink-like on microscopic scales, the string behaves as a free string
at long wavelength. The latter is the most elementary of all quantum
strings. It follows immediately that the relative transversal
displacement of two points separated by an arclength $l$ along the
string diverges as $\langle (z(l) - z(0))^2 \rangle \sim \ln
l$\cite{zhvs}.  The string as a whole is therefore delocalized, and
this is the simplest example of a ``critical'' string.

A central result of this paper is that Eq.\ (\ref{qusigo}) is, at
least in principle, not fully representative for the present lattice
problem. More precisely: starting from a more complete microscopic
kink dynamics model (section II) we find a richer infrared fixed point
structure. The phase diagram incorporates phases associated with the
quantum sine Gordon model fixed point, but also includes additional
phases which are intimately connected with the effects of the lattice
and of the nearest neighbor interactions between the holes. In section
III, we derive the path integral representation of our model. It turns
out that the worldsheet of this string in Euclidean space time
corresponds with two coupled restricted solid-on-solid (RSOS)
surfaces\cite{dennijsrev}, each of which describes the motion of the
string in either the $x$ or $y$ direction on the two dimensional
lattice.

The bare model is invariant under rotations of the string in space. As
discussed in section IV, we find indications for a generic
zero-temperature spontaneous symmetry breaking: for physical choices
of parameters, the invariance under symmetry operations of the lattice
is broken. Even when the string is critical (delocalized in space), it
acquires a sense of {\em direction}.  On average, the trajectories
corresponding with the string configurations move always forward in
one direction while the string might delocalize in the other
direction, see Fig.\ 1{\em f}. This involves an order-out-of-disorder
phenomenon which relatively easy to understand intuitively. Quantum 
mechanics effectively enhances the fluctuation dimension by stretching 
out the the string into a world sheet in the time-wise direction, and the 
enhancement of the effective dimension reduces the effect of fluctuations. 
Thermal fluctuations destroythis directedness, but they do so more 
effectively when the string isless quantum mechanical.

This directedness simplifies the remaining problem considerably.  We
will show that the directed string problem is equivalent to a well
known problem in surface statistical physics: its worldsheet is
equivalent to a single RSOS surface. At the same time, this model is
easily shown to be equivalent to a generalized XXZ quantum spin chain
problem. The particular model we study is actually equivalent to the
$S=1$ spin chain, which has been studied in great detail.  The RSOS
surface problem and the quantum spin chain problem are therefore also
related to each other. This equivalency was actually at the center of
the seminal work of den Nijs and Rommelse on the hidden order in
Haldane spin chains\cite{nijsrom}. From our perspective, the
introduction of the physically appealing quantum string model as an
intermediate model which connects both with the spin chain and the
RSOS surfaces, also helps to appreciate the depth of the work of den
Nijs and Rommelse\cite{nijsrom}.

The bulk of this paper (sections V-VIII) is devoted to an exhaustive
treatment of this directed string model.  Some powerfull statistical
physics notions apply directly to the present model, and these allow
us to arrive at a complete description of the phase diagram of the
quantum string. As was announced already in a short
communication\cite{eskes1}, this phase diagram is surprisingly rich:
there are in total ten distinct phases. In the context of the quantum
spin chain/RSOS surfaces, already six of those were previously
identified. However, viewing this problem from the perspective of the
quantum string, it becomes natural to consider a larger number of
potentially relevant operators and the other four phases become
obvious.

Compared to strings described by Eq.\ (\ref{qusigo}), we find a much
richer behavior but this is limited to the regime where lattice
commensuration dominates over the kinetic energy so that the string as
a whole is localized --- we use ``localized'' here in the sense that
the transversal string fluctuations of two widely separated points remain
finite, $\langle (z(\ell)-z(0))^2\rangle \rightarrow ~const.$ as $\ell
\rightarrow \infty$. Besides the different directions the purely
classical strings can take in the lattice, we also find a number of
localized strings which have a highly non-trivial internal structures:
the ``disordered flat'' strings, characterized by a proliferation of
kinks, but where the kink flavors condense so that the string as a
whole remains localized. On the other hand, the quantum-delocalized
(critical) strings are all of the free field variety and as we will
argue in the final section, this might be a very general consequence
of the presence of a lattice cut-off.

\section{Model: The meandering lattice string}\label{sec_string}

Whatever one thinks about the microscopy of the stripes, in the end
any theory will end up considering the charged domain walls as a
collection of particles bound to form a connected trajectory, or such
a model will be an important ingredient of it. Moreover, these
trajectories will communicate with the crystal lattice, because the
electrons from which the strings are built do so as well. This fact
alone puts some strong constraints on the collective dynamics of the
charged domain walls.

Let us consider the string configuration space. On the lattice this
will appear as a collection of particles living on lattice sites,
while every particle is connected to two other particles via links
connecting pairs of sites. The precise microscopic identity of these
particles is unimportant: they might be single holes (filled charged
domain walls\cite{zagu,harfo} as in the nickelates\cite{nickel}), an
electron-hole pair (the charge density waves of Nayak and
Wilczek\cite{nawi}, or Zaanen and Oles\cite{zaol}), or a piece of
metallic-\cite{metstr} or even superconducting\cite{whisca1} domain
wall. All what matters is that these entities have a preferred
position with regard to the underlying lattice (site
ordered\cite{zagu}, or bond ordered\cite{whisca}). Quite generally,
curvature will cost potential energy and a classical string will
therefore be straight, oriented along one of the high symmetry
directions of the lattice. Without loss of generality, it can be
assumed that the lattice is a square lattice while the string lies
along the (1,0) (`$x$') direction. Denoting as $N_y$ the number of
lattice sites in the $y$ direction and assuming periodic boundary
conditions, this straight string can be positioned in $N_y$ ways on
the lattice. Obviously, such a string will delocalize by {\em local}
quantum moves: the particles tunnel from site to
site\cite{prelov,vierrice}.  Moving the whole string one position in
the $y$ direction involves an infinity of local moves in the
thermodynamic limit, and the different classical strings occupy
dynamically disconnected regions of Hilbert space.

This is analogous to what is found in one dimensional systems with a
discrete order parameter\cite{revssh}.  In the case of, e.g.,
polyacetylene the order parameter is of the $Z_2$ kind: the bond order
wave can be either $\cdots -A-B-A-B- \cdots$ or $\cdots -B-A-B-A-
\cdots$ ($A$ single bond, $B$ double bond), while a single translation
over the lattice constant transforms the first state of the staggered
order parameter into the second kind of state. This is a discrete
operation, because the lattice forces the bond-order to localize on
the center of the bonds. Such an order parameter structure implies the
existence of topological defects, which are Ising domain walls:
$\cdots-A-B-A-B-B-A-B-A-\cdots$ (``kink'') and
$\cdots-B-A-B-A-A-B-A-B-\cdots$ (``antikink''). When they occur in
isolated form, these are also genuine building blocks for the quantum
dynamics, because although their energy is finite, it involves an
infinity of local moves to get rid of them (topological stability). In
the particular problem of polyacytelene, these kinks only proliferate
under doping (charged solitons). Although topological quantum numbers
are no longer strictly obeyed when the density of topological defects
is finite, it has been shown in a number of cases that they
nevertheless remain genuine ultraviolet quantities as long as they do
not overlap too strongly\cite{brazovski,haldane}.

If we consider a (locally) directed piece of string, the string is
analogous, except that the symmetry is now $Z_{N_y}$: on the torus, a
half-infinity of the string is localized at $y$ position $n_y$, and
the other half can be displaced to $n_y + 1, n_y + 2, \cdots n_y -1$.
Hence, in total there are $N_y - 1$ distinct kink excitations with the
topological invariants corresponding with the net displacement of the
half-string in the $y$ direction.  Because the kink operators can
occur in many flavors, this problem is therefore in principle richer
than that of one dimensional solids.

Clearly, kinks with different flavors have to be dynamically
inequivalent.  Since there is apparently a reason for the particles to
form connected trajectories, it should be more favorable to create a
kink corresponding with a small displacement than one corresponding
with a large jump. Here we will focus on the simplest possibility:
only kinks occur, corresponding with a displacement of {\em one}
lattice constant in the $y$-direction. This restriction is physically
motivated by the fact that the string is thought to separate two
antiferromagnetically ordered states; so, if the displacement of
successive holes would be larger than one lattice constant, the
antiferromagnetic ordering would be strongly suppressed --- after all,
this is the very reason that holes tend to line up in stripes.  In
addition, we will specialize on the ``neutral'' string. It will be
assumed that the string is characterized by a gap in its charge and
spin excitation spectrum, so that the strings with kinks contain the
same number of particles as the classical reference configurations.
The model we will consider might apply literally to the charge
commensurate stripes of the nickelate\cite{nickel}.  In the cuprates,
it might be better to consider the stripes as one dimensional metals
or superconductors, characterized by massless internal excitations. In
these cases, it remains to be demonstrated that eventually the
transversal string fluctuations decouple from the internal excitations
for the present model to be of relevance.
   
Given these considerations, we propose the following model for {\it
  quantum lattice strings}.  The string configurations are completely
specified by the positions of the particles (``holes'') ${\bf r}_l =
(x_{l},y_{l})$ on the 2D square lattice.  Two successive particles $l$
and $l+1$ can only be nearest or next-nearest neighbors, or $|{\bf
  r}_{l+1} - {\bf r}_l| = 1$ or $\sqrt{2}$.  We will call these
connections between successive particles {\it links}.  Two classes of
links, those of length $1$ and those of length $\sqrt{2}$, exist.
Taking the order of the particles into account there are $8$ distinct
links.  The string Hilbert space is spanned by all real space
configurations satisfying the above string constraint.

\begin{figure}
%\epsfxsize=1\hsize

%\vspace{0.5ex}
\epsfig{figure=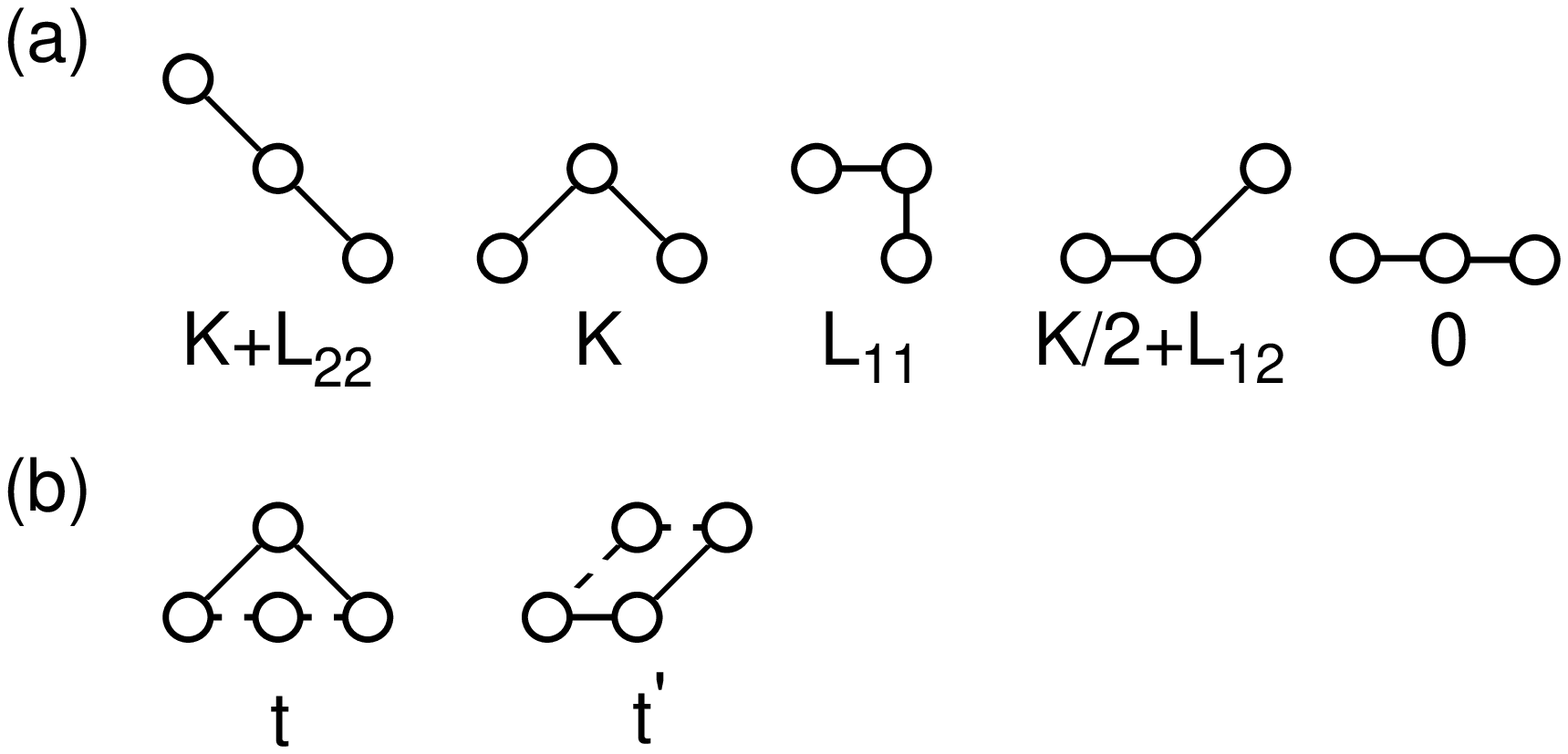,width=8cm}   
%\vspace{1.5ex}
\caption{ (a) The set of local configurations and their classical  
energies. (b) The two allowed hopping processes. We take  
$t = t^\prime$.  }
\label{fig_energies}
\end{figure}

We consider local discretized string-tension interactions between
nearest and next-nearest holes in the chain,(${\cal H} = -\beta H$)
\begin{eqnarray}
  {\cal H}_{Cl} &=& \sum_l \left[ {\cal K} \delta( | x_{l+1} - x_l | -
  1 ) \delta( | y_{l+1} - y_l | - 1 ) \right.  \nonumber \\ &+&
  \sum\limits_{i,j = 0}^{2} \left.  {\cal L}_{ij} \delta( | x_{l+1} -
  x_{l-1} | - i) \delta( | y_{l+1} - y_{l-1} | - j ) \right] \nonumber
  \\ &+& {\cal M} \sum\limits_{l,m} \delta( {\bf r}_{l} - {\bf r}_{m}
  )~.  \label{HCl}
\end{eqnarray}
The various local configurations and interaction energies are shown in
Fig.\ \ref{fig_energies}. The last term is an excluded volume type
interaction --- the physically relevant limit is $M \rightarrow
\infty$, so that holes cannot occupy the same site.  The interaction
${\cal K}$ distinguishes horizontal from diagonal links, and ${\cal
  L}_{ij} = {\cal L}_{ji}$ is a set of two-link interactions, which
one can think of as microscopic curvature terms.  Furthermore, we
exclude strings with a physically unrealistic extreme curvature by
taking ${\cal L}_{10}\rightarrow \infty$. Note also that
configurations which would give an contribution ${\cal L}_{00}$ to the
energy are automatically excluded in the limit $M\rightarrow \infty$,
which we will take throughout this paper. There are five local
configurations, distinguished by four parameters.  Therefore we can
choose ${\cal L}_{20}=0$ and the string is determined by the
parameters ${\cal K},{\cal L}_{11},{\cal L}_{12}$, and ${\cal
  L}_{22}$, see Fig.\ \ref{fig_energies}.

The second term in ${\cal H}$ is a quantum term which allows the
particles to hop to nearest neighbor lattice positions.  However such
hopping processes should not violate the string constraint.  This
constraint can be enforced by means of a projection operator
$P_{Str}({\bf r}_{l+1} - {\bf r}_{l} )$, that restricts the motion of
hole $l$ to the space of string configurations,
\begin{equation}
  P_{Str}({\bf r}) = \delta( |{\bf r}| - 1 ) + \delta( |{\bf r}| -
  \sqrt{2} )~.
\end{equation}
The string is quantized by introducing conjugate momenta
$\pi^{\alpha}_l$, $[ r^{\alpha}_l, \pi^{\beta}_{m} ] = i \delta_{l,m}
\delta_{\alpha,\beta}$, where $\alpha = x$ or $y$.  A term $e^{ i n
  \pi^{x}_l}$ acts like a ladder operator, and causes particle $l$ to
hop a distance $n$ in the $x$-direction,
\begin{equation}
  e^{ i n \pi^{x}_l} | x_{l} \rangle = | x_{l} + n \rangle.
\end{equation}
Therefore the kinetic energy term becomes,
\begin{equation}
  {\cal H}_{Q} = 2{\cal T} \sum\limits_{l,\alpha} P^\alpha_{Str}({\bf
    r}_{l+1} - {\bf r}_l) P^\alpha_{Str}({\bf r}_{l} - {\bf r}_{l-1})
  \cos(\pi^\alpha_l).  \label{HQ}
\end{equation}
Note that the particles, even under ${\cal H}_{Q}$, keep their order
and therefore can be labeled by $l$. Thus the fermion nature of holes 
in realistic domain walls plays no role at our level of approximation,
and quantum statistics becomes irrelevant.

The above model is minimal since it contains only nearest neighbor
hopping and the simplest string tension terms.  One natural extension
would be to take the two hopping amplitudes ${\cal T}$ in Fig.\ 
\ref{fig_energies}{\em b} to be different, since there is no microscopic
reason why they should be identical.  In the next sections we will
discuss the zero temperature properties of the above string model.
The self-avoidance term is a complicated non-local operator.  However,
we will find that, surprisingly, the kinetic energy favors {\it
  oriented} walls without loops.  Therefore this term turns out to be
unimportant for the present zero-temperature discussion.

\section{Relation to RSOS-like surface models} \label{sec_RSOS}

The problem introduced in the previous section can be reformulated as
the classical problem of a two dimensional surface (worldsheet)
embedded in 2+1 dimensional space, using the Suzuki-Trotter mapping.
The model can be seen as two coupled RSOS (restricted solid-on-solid)
surfaces.  The solid-on-solid models are classical models for surface
roughening\cite{dennijsrev}.  They describe stacks of atoms of integer
height in two dimensions, with an interaction between adjacent stacks
depending on the height differences.  With this construction overhangs
are excluded.  In the restricted SOS models these height differences
are limited to be smaller or equal to some integer $n$. In the present
case, the two $RSOS$ models parametrize the motions of the world sheet
in the spatial $x$- and $y$ directions, respectively, while the
(strong) couplings between the two takes care of the integrity of the
world sheet as a whole.

In the Suzuki-Trotter\cite{Suz76} or Feynman path integral picture one
writes the finite temperature partition function as an infinite
product over infinitesimal imaginary time slices.  In this limit the
commutators between the various terms in the Hamiltonian vanish like
${1/ n^2}$, where $n$ is the number of Trotter slices, and the
partition function can be written as,
\begin{equation}
  {\cal Z} = \lim_{n \rightarrow \infty} Tr \left( e^{ \frac{{\cal
        H}_{Cl}}{n} } e^{ \frac{{\cal H}_{Q}}{n} } \right)^n.
  \label{Z2}
\end{equation}
To show the relation with RSOS models, we will cast the transfer
matrices $T$ in the form of a two-dimensional classical effective
Hamiltonian.  This implies writing the matrix elements of the T-matrix
between configurations $\{ {\bf r}_{l} \}$ in terms of an effective
classical energy depending on the worldsheet positions $\{ {\bf
  r}_{l,k} \}$, where $k$ is the imaginary time index running from 1
to $n$ with periodic boundary conditions.  Schematically,
\begin{equation}
  \lim_{n \rightarrow \infty} \langle \{ {\bf r}_{l} \}_k |
  e^{\frac{1}{n}{\cal H}} | \{ {\bf r}_{l} \}_{k+1}\rangle \rightarrow
  e^{{\cal H}_{eff} (\{ {\bf r}_{l} \}_k, \{ {\bf r}_{l} \}_{k+1}) } .
                                                 \label{scematic}
\end{equation}
Since ${\cal H}_{Cl}$ is diagonal in the real-space string basis, it
is already in the required form,
\begin{equation}
  \lim_{n \rightarrow \infty} \langle \{ {\bf r}_{l} \}_k |
  e^{\frac{1}{n}{\cal H}_{Cl}} \rightarrow e^{\frac{1}{n}{\cal H}_{Cl}
    (\{ {\bf r}_{l} \}_k ) } \langle \{ {\bf r}_{l} \}_k | .
  \label{HCleff}
\end{equation}
For ${\cal H}_{Q}$ a few more steps are needed,
\begin{eqnarray}
  \lefteqn{ \langle \{ {\bf r}_{l} \}_k | e^{\frac{1}{n}{\cal H}_{Q}}
    | \{ {\bf r}_{l} \}_{k+1} \rangle} \hspace*{0.5cm} \nonumber \\ &
  & = \langle \{ {\bf r}_{l} \}_k | \sum\limits_{m=0}^\infty
  \frac{1}{m!} ( \frac{{\cal H}_{Q}}{n} )^m | \{ {\bf r}_{l} \}_{k+1}
  \rangle \nonumber \\ && = \langle \{ {\bf r}_{l} \}_k | 1 +
  \frac{{\cal H}_{Q}}{n} | \{ {\bf r}_{l} \}_{k+1} \rangle ~+~ {\cal
    O}({{1}\over{n^2}}) \nonumber \\ && = \prod\limits_l
  \prod\limits_{\alpha=x,y} ( \delta ( {\alpha}_{l, k+1} -
  {\alpha}_{l,k} ) \nonumber \\ && \hspace{3cm} +{{\cal T} \over n}
  \delta ( | {\alpha}_{l, k+1} - {\alpha}_{l,k} | - 1 )) \nonumber \\ 
  && = e^{ \sum_l \ln \left( { {\cal T} \over n} \right) \left[ \delta
    (| x_{l, k+1} - x_{l,k} | - 1 ) + \delta (| y_{l, k+1} - y_{l,k} |
    - 1 ) \right] }.
                                                  \label{HQeff}
\end{eqnarray}
The expression in the last line is of course only valid for states in
which the $\alpha_l$'s in successive time slices differ by at most
one unit.  Combining these two energy contributions we arrive at the
following classical problem,
\begin{eqnarray}
  && {\cal Z} = \lim_{n \rightarrow \infty} Tr e^{ {\cal H}_{eff} }
  \nonumber \\ &&{\cal H}_{eff} = \sum_{l,k} \left[ { {\cal K} \over
    n} \delta( |x_{l+1,k}-x_{l,k}| - 1 ) \delta( |y_{l+1,k}-y_{l,k}| -
  1 ) \right.  \nonumber \\ && + \sum\limits_{i,j = 0}^{2} {{\cal
      L}_{ij} \over n} \delta( | x_{l+1,k} - x_{l-1,k} | - i) \delta(
  | y_{l+1,k} - y_{l-1,k} | - j ) \nonumber \\ && + {M \over n}
  \sum\limits_{m} \delta( x_{l,k} - x_{m,k} ) \delta( y_{l,k} -
  y_{m,k} ) \nonumber \\ && + \ln ( {{\cal T} \over n} ) \left[ \delta
  ( | x_{l,k+1} - x_{l,k} | -1 ) \right.  \nonumber \\ && \hspace{1cm}
  \left.  + \delta ( | y_{l,k+1} - y_{l,k} | -1 ) \right].
    \label{Heff}
\end{eqnarray}
This classical world sheet is constrained to $| x_{l,k+1} - x_{l,k} |
\leq 1$ and $| y_{l,k+1} - y_{l,k} | \leq 1$, and the interactions are
anisotropic.  The above classical model can be viewed as two coupled
two-dimensional RSOS surfaces, $x_{l,k}$ and $y_{l,k}$.  The $x$
coordinate of hole $l$ at the time slice $k$ is now identified as the
height of an RSOS column positioned at $(l,k)$ in the square lattice.
In a similar way the $y$ coordinates define a second RSOS surface,
coupled strongly to the first by the above classical interactions.
Since the steps $\Delta x$ can at most be equal to $1$, the RSOS
sheets are restricted to height differences $0$, $\pm 1$ between
neighboring columns.

The classical model as defined above is not unique.  While the above
mapping allows us to exploit the connection to other models most
efficiently, for the numerical Monte-Carlo calculations a different
decomposition is used, which allows for a more efficient approach to
the time continuum limit. We write,
\begin{eqnarray}
  {\cal Z} & = & Tr e^{ {\cal H}_{Cl} + {\cal H}_{Q} } \nonumber \\ &
  = & \lim_{n \rightarrow \infty} Tr \left( I T_{A} I T_{B}
\right)^{n} .  \label{Z}
\end{eqnarray}
In the above formula $I$ is the identity operator, in our case a
complete set of string configurations.  We have chosen to split the
$T$-matrix into a contribution from even and odd sites, or A and B
sublattice (checkerboard decomposition),
\begin{equation}
  T_{A} = \exp\left[ \frac{1}{n} \sum\limits_{l=1}^{L/2} \left( {\cal
    H}_{Cl,2l} + {\cal H}_{Q,2l} \right) \right], \label{T_A}
\end{equation}
and a similar expression for the odd sites.  ${\cal H}(2l)$ is the
Hamiltonian of the even string element $2l$, equal to Eq.\ (\ref{HCl})
or (\ref{HQ}) without the sum over string links.  $L$ is the number of
links in the chain.  Because of the sublattice decomposition, $T_{A}$
is a simple product of local T-matrices and ${\cal Z}$ becomes,
\begin{eqnarray}
  {\cal Z} = \lim_{n \rightarrow \infty} \sum\limits_{ \{ {\bf
      r}_{l,k}, {\bf r}^{\prime}_{l,k} \} } \prod\limits_{k=1}^{n}
  \prod\limits_{l=1}^{L/2} && \langle \{ {\bf r}_{l} \}_k |
  t^{2l,k}_{A} | \{ {\bf r}_{l} \}^\prime_k \rangle \nonumber \\ &&
  \langle \{ {\bf r}_{l} \}^\prime_k | t^{2l+1,k}_{B} | \{ {\bf r}_{l}
  \}_{k+1} \rangle .
                                                         \label{Zt}
\end{eqnarray}
Each timeslice is split in two subslices, ${\bf r}$ and ${\bf
  r}^\prime$.  The notation $\{ {\bf r}_{l} \}_k$ denotes the set of
positions $ {\bf r}_{l}$ at the given time slice with index $k$.  Note
that the $t$ matrices are independent of $l$ and $k$ , and these
indices only label the position of the $t$ matrix in the 2D world
sheet.  The local $t$ matrices, $t_A$ and $t_B$, depend only on three
positions.  For instance,
\begin{eqnarray}
  && \langle \{ {\bf r}_{l} \}_k | t^{2l,k}_{A} | \{ {\bf r}_{l}
  \}^\prime_k \rangle \nonumber \\ && = \langle {\bf r}_{2l-1,k} {\bf
    r}_{2l,k} {\bf r}_{2l+1,k} | t^{2l,k}_{A} | {\bf
    r}_{2l-1,k}^\prime {\bf r}_{2l,k}^\prime {\bf r}_{2l+1,k}^\prime
  \rangle
                                                         \label{t6}
\end{eqnarray}
with the restriction ${\bf r}_{2l-1,k} = {\bf r}_{2l-1,k}^\prime$ and
${\bf r}_{2l+1,k} = {\bf r}_{2l+1,k}^\prime$.  Since each link has 8
different orientations, the local $t$ matrix connects in general 
$8\times8 = 64$ possibilities.  However most of the $t$ matrix elements 
are zero, and it decomposes into subblocks, of which the biggest one is
$3\times 3$.  The states which are connected via the local $t$ matrix,
or the Hamiltonian, are listed in Fig.\ \ref{fig_Tblocks},  and the
matrix elements are given in the appendix.  The above decomposition
will be used in the Monte-Carlo simulations described in Section
\ref{sec_full}.
\begin{figure}
%\vspace{0.5ex}
\epsfig{figure=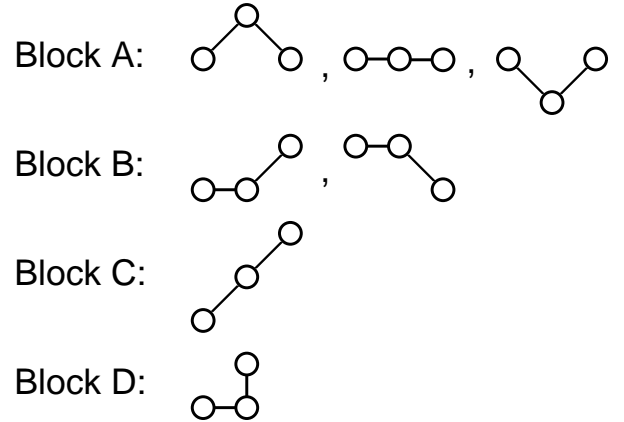,width=8cm}   
%\vspace{1.5ex}
\caption{ The four subblocks of the local $t$-matrix.  The other 
equivalent, symmetry related blocks are obtained by $\pi/2$ 
rotations and reflections in the $x$ or $y$ axis.  }
\label{fig_Tblocks}
\end{figure}

\section{Directedness as spontaneous symmetry breaking}\label{sec_full}

We are not aware of any similarity of the statistical physics problem
of the previous section with any existing model. RSOS problems are
well understood, but it should be realized that in the present model
the two RSOS problems are strongly coupled, defining a novel dynamical
problem.  When we studied this problem with quantum Monte-Carlo, we
found a generic zero temperature symmetry breaking: although the
string can be quantum delocalized, it picks spontaneously a {\em
  direction} in space. This symmetry breaking happens always in the
part of parameter space which is of physical relevance. Our 
understanding of this phenomenon is a nice illustration of an 
order-out-of-disorder mechanism which at first sight appears paradoxical, 
but which is easy to understand qualitatively. At first sight, one
might expect that the quantum fluctuations (kinetic energy) would tend
to disorder a string, i.e., to decrease the tendency for the string to
be directed. That the opposite effect happens, is due to the fact that
in the presence of quantum flucuations, the string traces out a
worldsheet in the imaginary time direction. The larger the kinetic
term, or the larger the temperature, the further the worldsheet
stretches out in the timewise direction. At zero temperature, the
worldsheet becomes infinite in this direction as well. The statistical
physics of a string is then equivalent to that of a fluctuating sheet
in three dimensions. Now, it is well known from studies of classical
interfaces\cite{wee77} that while a one-dimensional classical
interface in two dimensions does not stay directed due to the strong
fluctuations, for a two-dimensional sheet the entropic fluctuations
are so small that interfaces can stay macroscopically flat in the
presence of a lattice\cite{Wee80,Bei87} --- for this reason, the
roughening transition in a three-dimensional Ising model is properly
described by (i.e., is in the same universality class as) a
Solid-on-Solid model in which overhangs are
neglected\cite{Wee80,Bei87}. In other words, even if microscopic
configurations with overhangs are allowed, a classical interface on a
lattice in three dimensions can stay macroscopically flat or
``directed''. In agreement with this picture, we will find that the
directedness order becomes more robust when the string
quantum-fluctuations become more severe, i.e., when the wordsheet
stretches out further into the timewise direction.

It is convenient to use open spatial boundary conditions. In this
case, the string configuration space of the model defined in section
II consists of all surfaces, connecting to the boundaries in the
space-directions while periodic along the imaginary time direction. On
a time slice, the trajectories can be categorized according to the
global property of {\em directedness}: when the string connects two
opposite sides of the lattice in the spatial directions, this
trajectory can be called ``directed'' and otherwise it is undirected.
An alternative, more restrictive definition of directedness is: every
continuous string configuration $s$ can be written as a parametrized
curve in two dimensions, [$x(t),y(t)$], where $t$ could for instance
be the discrete label of the successive particles along the string.
When the string configuration can be parametrized by a single-valued
function $x(y)$ or $y(x)$, we call the string configuration directed
(see Fig.\ \ref{fig_walls}{\em f}). The quantum string vacuum is a
linear superposition of many string configurations. When all
configurations in the vacuum correspond with single valued functions
$x(y)$ or $y(x)$, the string vacuum has a directedness order
parameter. We will give an explicit measure of the directedness in our
simulations below.

It should be immediately clear that directedness order is rather
fragile.  {\em It cannot exist at any finite temperature.} When
temperature is finite, the width of the worldsheet in the imaginary
time direction becomes finite as well, and the long wavelength
fluctuations of the string becomes a 1D statistical problem, subjected
to string boundary conditions which are inconsequential for the
argument which follows.  Consider first the classical limit: at zero
temperature the string would be straight, running along (say) a (1,0)
direction. A local `corner' configuration of the type shown in Fig.\ 
\ref{fig_energies}{\em a} would be an excitation with energy $L_{11}$
(alternatively, one could consider two kinks). Clearly, a single
corner suffices to destroy the directedness of the classical ground
state.  At any finite temperature, the probability of the occurrence
of at least one corner is finite: $P = N \exp ( - \beta L_{11})$.
Hence, directedness order cannot exist at finite temperatures, for the
same reasons that long range order is destroyed at any non-zero temperature 
in one dimension. 

\begin{figure}
%\vspace{0.5ex}
\epsfig{figure=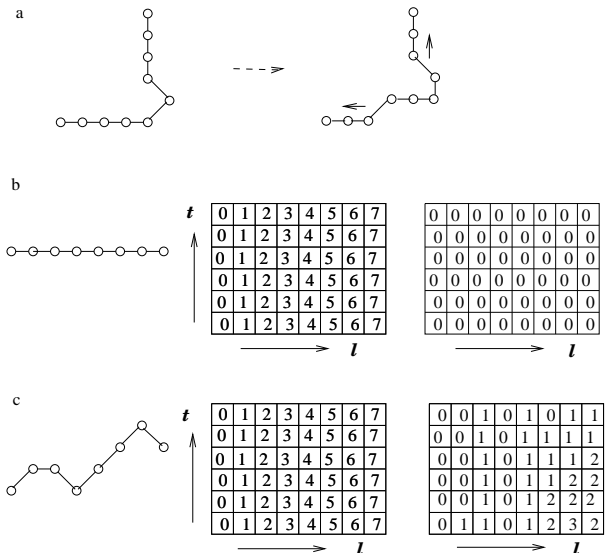,width=8cm}   
%\vspace{1.5ex}
\caption{ (a) An undirected string with two kinks propagating 
along different directions. Note that the bend blocks the propagation 
of kinks. (b) a (1,0) string and the corresponding two (coupled) 
RSOS surfaces along the $x$ and the $y$ directions respectively. 
The numbers  correspond to $x$ ($y$) position of hole $l$ at imaginary 
 time $t$. (c) a disordered {\em directed} string and the corresponding 
 ordered and  disordered RSOS surfaces }
\label{figure_blocks}
\end{figure}

At zero temperature, the internal dynamics of the worldsheet is that
of a two-dimensional sheet which is infinite in both directions and
since the lattice commensuration has rendered the problem to be of a
discrete symmetry, long range order can exist for finite kinetic
energy. We will discuss the symmetry breakings of a single RSOS
surface in great detail later. For the present discussion it suffices
to know that such a single RSOS surface can be fully ordered, as well
as (partly) disordered. Because of the strong coupling, it would a
priori appear questionable to discuss the dynamics of the full model
of Section II in terms of the dynamics of the two separate RSOS
subproblems. However, in the context of directedness it is quite
convenient to do so. When both the $x$ and $y$ RSOS problems would be
fully disordered, it is easy to see that the string vacuum would be
undirected. This is illustrated in Fig.\ \ref{figure_blocks}{\em a}:
two kinks moving the  string from a (1,0) to a (0,1) direction in the
lattice correspond with one kink which can move freely in the
horizontal part of the string, and one kink which can move freely in
the vertical part of the string. On the other hand, when both RSOS
problems are ordered, the string is also ordered. For instance, the
(1,0) string can be thought of as a combination of an RSOS surface
which always steps upwards in the $x$ direction, and one which is
horizontal in the $y$ direction (Fig.\ \ref{figure_blocks}{\em b}).

A third possibility is that one of the RSOS subproblems is ordered,
while the other is disordered. Dismissing crumpled phases (like
condensates of the $L_{11}$ type corners), the only possibility
remaining is that one of the RSOS problems steps up always, while the
other is disordered, as illustrated in Fig.\ \ref{figure_blocks}{\em c}. 
This results in a disordered {\em directed} string vacuum: the string steps
always forward in, say, the $x$ direction while it freely fluctuates
in the $y$ direction. We find that this symmetry breaking happens {\em
  always} as long as the string does not crumple.

\begin{figure}
%\epsfxsize=0.8\hsize

%\vspace{0.5ex}
\hspace{-2ex}\epsfig{figure=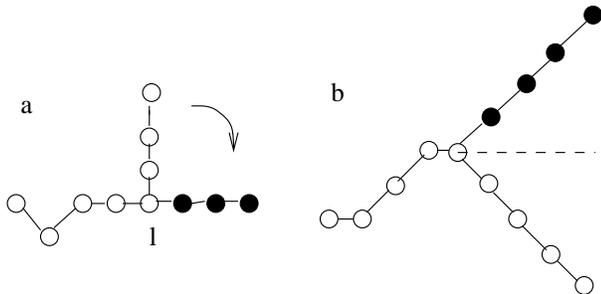,width=8cm}   
%\vspace{1.5ex}
\caption{Two additional Monte Carlo operations used for 
the simulations of the general string. The 90 degree rotation 
(a) around position l in this example turns a non-directed 
string into  a directed one. The mirror operation (b) is 
important to quench defects in diagonal strings (mirror plane 
indicated by the dashed line).}
\label{figure_mcglobal}
\end{figure}

To perform Monte-Carlo calculations on the string model, we used the
checkerboard decomposition as described in Sec.\ \ref{sec_RSOS}.  Note
that the string has no minus sign problem --- the sign of ${\cal T}$
is easily transformed away by an appropriate choice of the operators.
For ${\cal L}_{10}\rightarrow \infty$, there are 8$\times$5=40
different local configurations, given by the positions ${\bf
  r}_{l-1}$, ${\bf r}_{l}$ and ${\bf r}_{l+1}$.  However most of the
$40\times 40$ local $T$ matrix elements are zero, and the largest
subblock is $3\times 3$.  These blocks were listed in Fig.\ 
\ref{fig_Tblocks}.  The corresponding matrix elements are listed in the
appendix.  For the Monte-Carlo program to produce sensible results it
is crucial to have operations that add and remove bends easily.  We
added global mirror and $\pi/2$ rotation operations illustrated in
Fig.\ \ref{figure_mcglobal}.
 In the latter case half of the string is
rotated around any of the sites $l \in 2 ..  L-1$.  This means that
for instance the position of all holes $m > l$ are replaced by
$(x_m,y_m) \rightarrow (x_l,y_l) + (y_m-y_l,-(x_m-x_l))$.  Such
operations turn out to be very efficient --- completely wrapped high
temperature strings unwrap in just a couple of Monte Carlo steps at
low temperature.

We discovered this directedness symmetry breaking during the
simulations.  To quantify the directedness we measured the fraction of
string length in the string vacuum corresponding with single valued
configurations. At zero temperature, the ground state wave function of
the string is,
\begin{equation} \label{wavefun}
  | \Psi_0 \rangle = \sum_{\{ x_l, y_l \} } \alpha_0 ( \{ x_l, y_l \})
  | \{ x_l, y_l \} \rangle~.
\end{equation}
where every state in string configuration space ($| \{ x_l, y_l \}
\rangle$) corresponds with a trajectory [$x(t),y(t)$]. Consider first
the case of a continuous string. For every configuration, the total
string arclength is given by
\begin{equation} \label{totl}
  L(\{x_l,y_l\})_{tot}= \int ds = \int \sqrt{dx^2+dy^2}~.
\end{equation} 
\begin{figure}
%\epsfxsize=0.8\hsize

%\vspace{0.5ex}
\hspace{3ex}\epsfig{figure=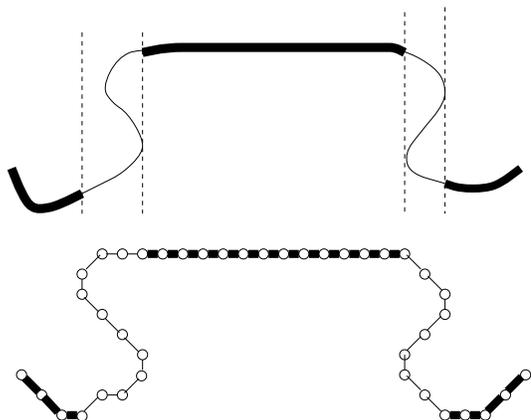,width=7cm}   
%\vspace{1.5ex}
\caption{ Illustration of the way we measure the directedness 
of a string in the continuum case (a) and on the lattice (b). 
The heavy solid parts of the string  indicate the parts where the 
projection of the string onto the $x$ axis is single-valued, and for 
which the indicator function $g_y(x)$ equals 1. }
\label{directeddef}
\end{figure}
To quantify the directedness, we introduce, as illustrated in Fig.\ 
\ref{directeddef}, an indicator function $g_y(x)$ which equals 1 when
the string is single-valued when projected onto the $x$-axis, and zero
otherwise, and analogously a function $g_x(y)$ which equals 1 when the
curve is single-valued when projected onto the $y$-axis, and zero
otherwise. The total directed length in the $x$ and $y$ directions are
then defined as
\begin{eqnarray} \nonumber
  L(\{ x_l, y_l \})_{dir, x} & = & \int dx ~ g_y(x) \sqrt{1 + \left(
    {{dy}\over{dx}}\right)^{2} }~,\\ L(\{ x_l, y_l \})_{dir, y} & = &
  \int dy ~ g_x(y) \sqrt{1 + \left( {{dx}\over{dy}}\right)^{2}
    }~.\label{totdx}
\end{eqnarray}
The measure of directedness in the string vacuum is then defined as
the larger of $N_{dir}^{x} (0)$ and $N_{dir}^y (0)$, where
\begin{equation} \label{dirden}
  N_{dir}^{\eta} ( 0 ) = \sum_{ \{ x_l, y_l \} } | \alpha^0 ( \{ x_l,
  y_l \} )|^2 { {L(\{ x_l, y_l \})_{dir, \eta}} \over { L({x_l,
        y_l})_{tot} } }~,
\end{equation}
where $\eta=x,y$. On the lattice, our measure of directedness is the
immediate analogue of this definition, except that we just count the
number of directed bonds, irrespective of whether they are oriented
diagonally or horizontally.

By thermal averaging, the above definition of directedness density is
immediately extended to finite temperature,
\begin{equation} \label{dirdenT}
  N^{\eta}_{dir} ( T ) = \sum_{n} e^{ -\beta (E_n - E_0) }
  N_{dir}^{\eta} ( n )~,
\end{equation}  
where $N_{dir}^{\eta} ( n )$ is the directedness density of a string
excitation with energy $E_n$. Eq. (\ref{dirdenT}) can be
straightforwardly calculated using quantum Monte-Carlo. A Monte-Carlo
snapshot defines a stack of coupled string configurations along the
imaginary time direction (the Trotter direction). We calculate
$N_{dir}^{x} $ for every Trotter slice by calculating the fraction of
the string length in this configuration which is single valued in the
$x$-direction. This is given by the number of bonds which steps
forward in the $x$-direction devided by the total number of bonds in
the string. We then average this quantity over the string wold sheet
(Trotter direction) and then over the Monte-Carlo measurements. The
same is done for $N_{dir}^{y} ( n )$. The larger of $N_{dir}^{x} ( n
)$ and $N_{dir}^{y} ( n )$ is then the density of directedness at the
given temperature.

\begin{figure}
%\epsfxsize=1\hsize

%\vspace{0.5ex}
\epsfig{figure=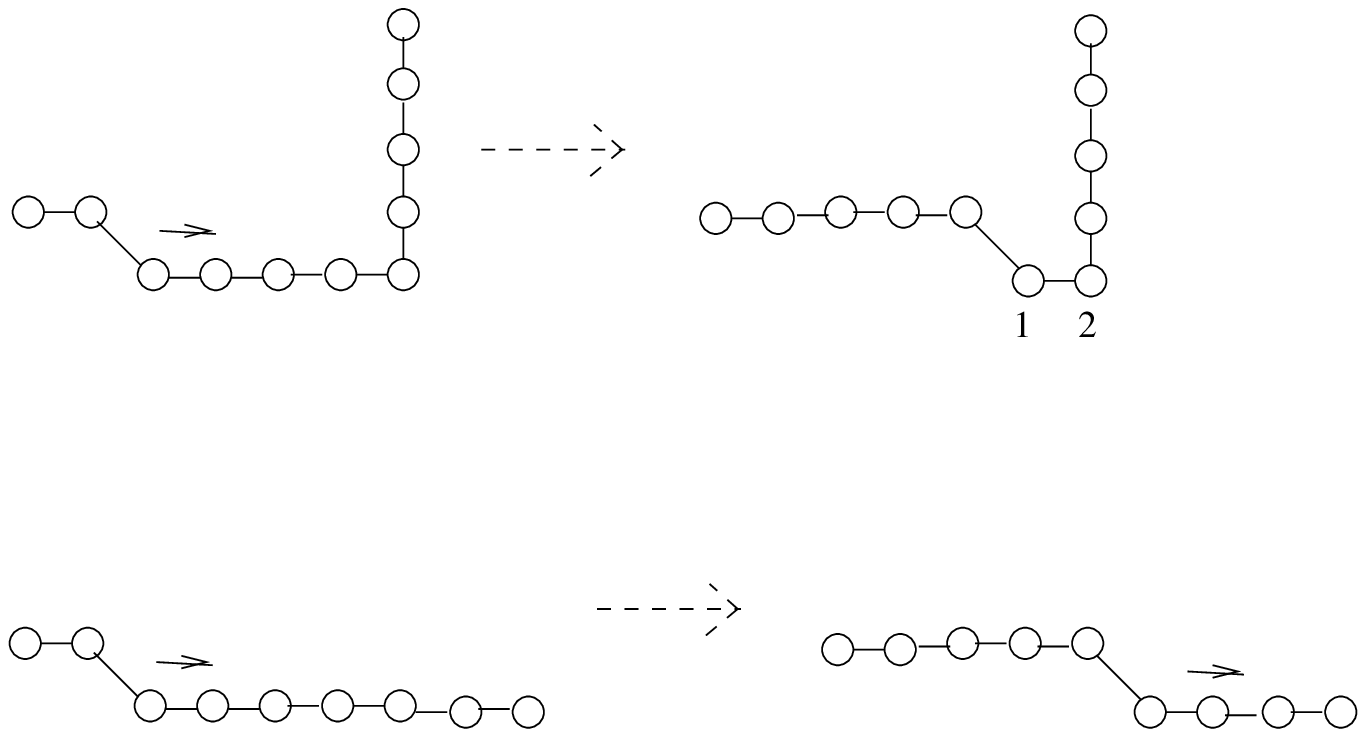,width=8cm}   
%\vspace{1.5ex}
\caption{ Illustration of the fact that a bend blocks the 
propagation of links along the string. Note holes 1 and 2 cannot 
move.} 
%\protect\label{fig_bends}
\label{fig_bends}
\end{figure}

That the zero-temperature properties of the directed and non-directed
string are equivalent for ${\cal L}_{11} \ge 0$ can be understood by
the following argument.  The $\pi/2$ bends in strings block the
propagation of links along the chain.  Close to the bend itself the
particles in the chain cannot move as freely as in the rest of the
chain. This effect is shown in Fig.\ \ref{fig_bends}.
 In space-time
the $\pi/2$ bend is like a staight rod in time.  Therefore the
presence of such kinks increase the kinetic energy.  For the argument
it makes little difference whether the bend consists of a single
$\pi/2$ corner or two $\pi/4$ corners.  This confirms that {\it it is
  the kinetic energy which keeps the strings oriented along one
  particular direction}.  In terms of a directedness order parameter
this result implies that such a quantity is always finite, except when
${\cal L}_{11} \ll 0$ or when the hopping term vanishes (it is easy to
see that in the classical case, ${\cal T}=0$, in many regions of
parameter space the problem becomes that of a self-avoiding walk on a
lattice in the limit $T\rightarrow 0$). For the two equivalent RSOS
surfaces this means that one of the two spontaneously orders and
becomes diagonal, while the other RSOS sheet shows any of the ten
quantum rough or classically ordered phases discussed later.

\begin{figure}
\epsfxsize=0.9\hsize

%\vspace{0.5ex}
\hspace{-3ex}\epsfig{figure=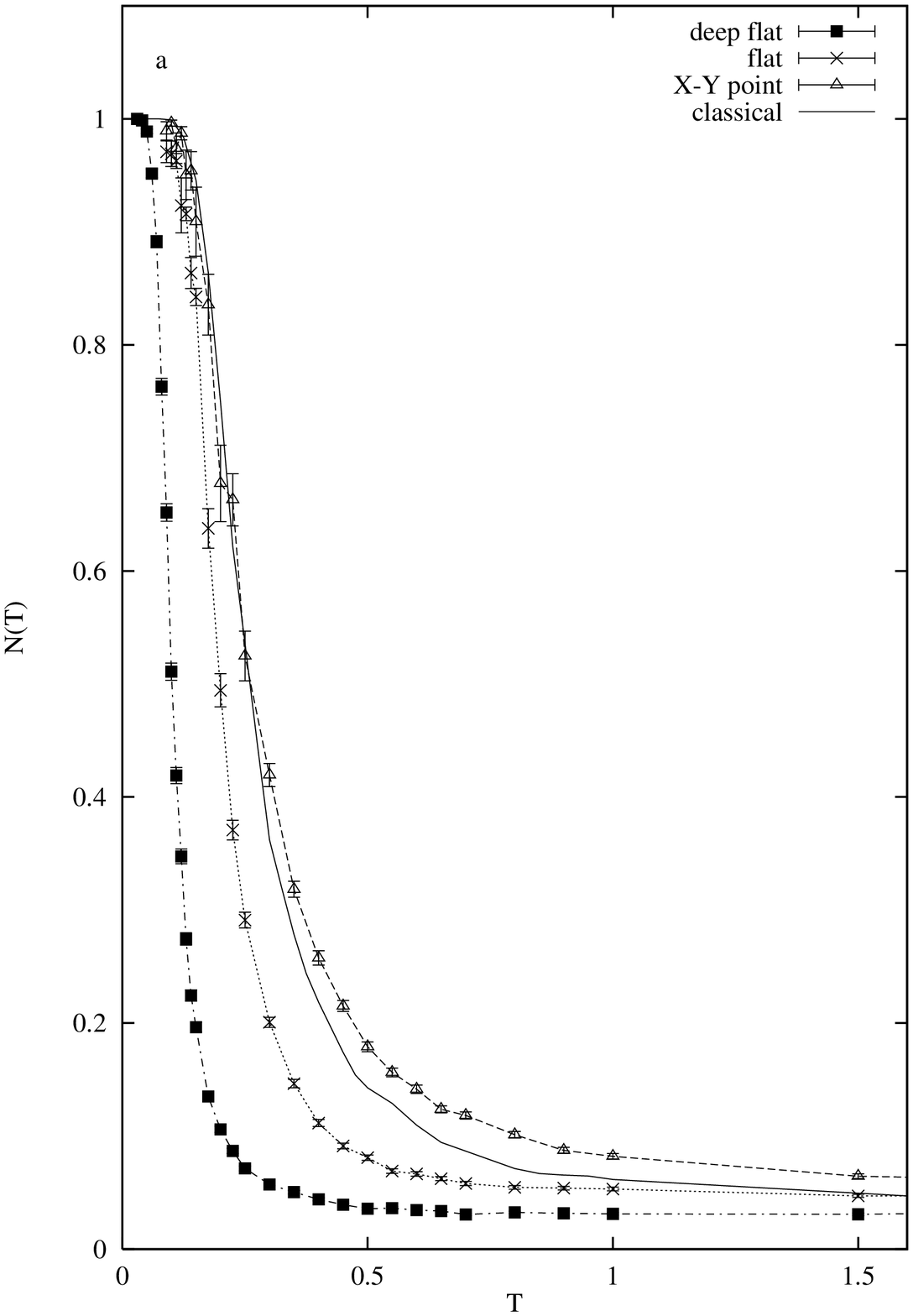,width=\linewidth}   
%\vspace{-70ex}
%\hspace{15ex}\epsfig{figure=dir.ord.b.ps,width=4cm}   
\caption{ A Monte-Carlo result for the directedness density $N_{dir} ( T )$ 
at 4 points. (a) The XY point (Triangles) where all curvature energie 
are zero. Two points in the flat phase, with ${\cal K} = 1.8$ (crosses) 
and ${\cal K} = 4.0$ (filled squares) the rest of the curvature energies 
are zero. (b) Inset: A point in the middle of the gaussian phase with 
parameters ${\cal K} = 0.5$, ${\cal L}_{21} = -0.25$, ${\cal L}_{22} = -1.0$, 
${\cal L}_{11} = 0$ (open circles). The full line in both figures is the 
result for a classical string where only flat bonds and ${\pi \over 2}$ 
corners are present with ${\cal L}_{11}=1$. }

%\vspace{0.5ex}
%\hspace{5ex}\epsfig{figure=dir.ord.eps,width=\linewidth}   
%\epsfig{figure=dir.ord.b.ps,width=\linewidth}   
%\vspace{1.5ex}
\label{fig_dir_ord}
\end{figure}

In Fig.\ \ref{fig_dir_ord} we show results of typical Monte-Carlo
calculations for the density of directedness as a function of
temperature $N_{dir} ( T )$.  We have considered four points in the
parameters space; as will be discussed later, these points are
representative for phases with interesting quantum fluctuations, and  
serve to clarify our conclusion. In Fig.\ \ref{fig_dir_ord}(a) the triangles 
(dashed line) is the result for the density of directedness at the point 
where all the classical curvature energies are zero, i.e., corresponding to 
the pure quantum string. The crosses (dotted line) and the filled squares 
(dashed-dotted line) are the results for points where ${\cal K} = 1.8$ and 
${\cal K} = 4.0$ respectively, and the rest of the classical curvature 
energies are zero. In term of the phase diagram for the directed string 
problem of Fig.\ \ref{fig_pd0} in section VI and Table \ref{tabPhases} in 
section VII, the first point corresponds to a Gaussian string (pure quantum) 
and the other two correspond to flat strings. The point ${\cal K} = 1.8$ 
lies just inside the flat string phase II where significant quantum 
fluctuations are still present, while the point ${\cal K} = 4.0$ lies 
deep inside the flat phase. The fourth curve in Fig.\ \ref{fig_dir_ord}(a), 
given by the full line, is the result of a Monte-Carlo calculation for a 
classical string (${\cal T}=0$) where only flat segments and ${\pi/2}$ 
corners are allowed (no diagonal segments). This same classical result is 
shown again in Fig.\ \ref{fig_dir_ord}(b) together with the result of the 
directedness density for a point in the midlle of the Gaussian (XY) phase 
[${\cal K} = 0.5$, ${\cal L}_{21} = -0.25$, ${\cal L}_{22} = -1.0$, 
${\cal L}_{11} = 0$ corresponding to $D =0$ and $J = -0.5$].

We will first discuss the result for the classical case, a classical
string (${\cal T}=0$) with ${\cal L}_{11}$, the energy of the ${\pi/2}$ 
corner, equal to one.  Because the string is of finite length, the 
infinite temperature limit of $N_{dir} ( T )$ is not zero but rather a 
small but nonzero value\cite{selfavoid} ($\sim 0.03$ for a domain wall 
of length 50).  $N_{dir} ( T )$ is already close to this value for all 
temperatures of order ${\cal L}_{11}$ and larger. For an infinitely long 
domain wall $N_ {dir}( T )$ drops very fast to zero with increasing 
temperature. At the other limit, for low $T$ where $T \ll {\cal L}_{11}$, 
$N_{dir} ( T )$ grows very fast to 1.

\vskip -100ex
\begin{figure}[t]
%\epsfxsize=0.5\hsize

%\vspace{-70ex}
\hspace{17ex}\epsfig{figure=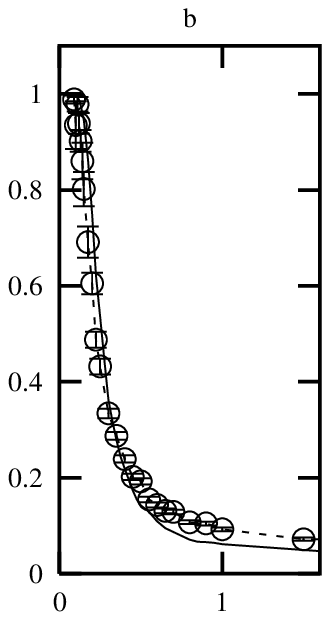,width=5cm}   
%\vspace{30ex}
\end{figure}
%\vskip 90ex
\begin{figure}
\begin{picture}(20,168)
\end{picture}
\end{figure}
%\newpage
%\noindent

  Again because the string is of finite 
length, it becomes directed already at a finite temperature: For all 
temperatures such that $L\exp(-{\beta}{\cal L}_{11}) < 1$ the string 
configurations in our simulations are typically completely directed. An 
infinitely long classical string becomes directed only at $T = 0$, of 
course, since at any nonzero temperature always some corners will occur 
in a sufficiently long string. 

For the quantum string, all the curves look strikingly similar to the
classical one. When the temperature is very much higher than the
kinetic term, $T \gg {\cal T}$, all curves merge together and the
classical limit is reached. At low $T$, where $T \ll {\cal T}$,
$N_{dir} ( T )$ again grows very fast to 1, as in the classical case;
it reaches this value at a finite temperature for the finite length
string. This is even true for the purely quantum string at the XY
point, where all classical microscopic curvature energies are zero
(see the dashed line in Fig.\ \ref{fig_dir_ord}(a)). We can understand 
this in terms of an effective corner or bend energy ${\bar {\cal L}}$ 
that is produced by the quantum fluctuations. As in the classical case
the probability for the occurance of a bend is proportional to 
$\sim \exp(- \beta {\bar {\cal L}})$.  At zero temperature no bend 
is present and the string becomes directed. A finite length string 
effectively becomes directed already at a temperature such that 
$L \exp ( - \beta{\bar {\cal L}}) < 1$. At intermediate temperatures, 
where the temperature is of the order of the kinetic term, things are 
more difficult and it is far from obvious what is going on. Especially 
in this region, all the various classical curvature energies may play a
role, and the interplay of these on the directedness is unclear.
Nevertheless, as is clear from the data of Fig.\ \ref{fig_dir_ord}(a),
this region connects the high and low temperature limits smoothly.
Moreover, by comparing the results for the three quantum strings in this 
figure it is also clear that when the string is more quantum mechanical 
$ N_{dir} ( T )$ is higher.

Our general conclusion, based also on Monte-Carlo studies of the 
behavior in other phases which are briefly summarized below, is that
{\it apart from some extreme classical limits, the general lattice string model at zero 
temperatures is a directed string}. The qualitative picture of $\pi/2$ 
turns blocking the propagation of kinks appears to
be a natural explanation for these Monte-Carlo findings. The phase
diagram of the general string model introduced in section II will
essentially be the same (apart from special limits) as the
corresponding phase diagram of the simplified directed string model.
In the remaining sections of this paper we will therefore focus on the
phases and phase transitions of the directed string.

\begin{figure}
%\vspace{0.5ex}
\hspace{7ex}\epsfig{figure=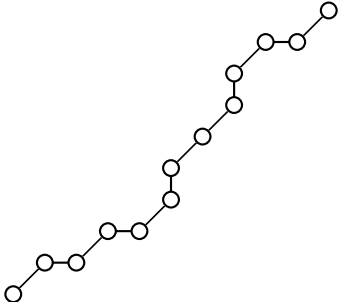,width=4.5cm}   
%\vspace{1.5ex}
\caption{ A typical low temperature string in the slanted parameter  
region VII.}
%\protect\label{fig_nondirslanted}
\label{fig_nondirslanted}
\end{figure}

For completeness, we end this section with a brief qualitative
description of our observations concerning spontaneous directedness at
low but finite temperatures in regions of the phase diagram where the
directed string has other type of ordering than that already
discussed.  All the results apply to ${\cal L}_{11}=0$, and we refer
to Table \ref{tabPhases} in section VII for a quick introduction to
the various 
 phases of the directed string problem and for the
 numbering (I--X) of the varous phases. \\  --- The entire zero
temperature phase diagram of the directed string is reproduced.\\ ---
Phase I is very stable with respect to bends. With ``stable'' we mean
that {\it finite} strings do not change their appearance when
increasing the temperature from zero to a moderately small
temperature, of the order of 0.1 ${\cal T}$.\\ --- Deep in the
horizontal phase II (large positive ${\cal K}$) quantum fluctuations
are strongly suppressed, and at the same time the string becomes
susceptible to $\pi/2$ corners. On the other hand, when we approach
from phase II the boundaries with phases IV and V, the fluctuations
increase and the string stiffens, Fig.\ \ref{fig_dir_ord}.  
This is in agreement with the picture sketched before that quantum 
fluctuations orient the string.\\ 
--- Deep inside phase III the string changes constantly between
horizontal zigzags and vertical zigzags.  A $\pi/2$ turn costs no
extra energy.  Again close to phase V quantum fluctuations have the
effect of removing bends.\\ --- The Haldane phase V and the rough
phase IV are very robust, and a considerable fraction of $\pi/2$ bends
occurs only at relatively high temperatures of the order of 0.2 ${\cal
  T}$.\\ --- In the slanted phase VII high temperatures are needed
before down diagonal links come in.  On the other hand horizontal
links are easily replaced by vertical ones.  This only increases the
energy very slightly, but the entropy gain is considerable.  A typical
low temperature string is shown in Fig.\ \ref{fig_nondirslanted}
 in section VII.  To zeroth order the horizontal and vertical links can be
thought of as spinless fermions moving coherently along the string.
In the dilute limit these links have only a weak interaction.  The
order of the links is conserved, and at zero temperature the ground
state has only horizontal links.  However our simulations indicate
that for a small range of negative $L_{11}$ values a diagonal string
with alternating horizontal and vertical links is favoured.  It is
again the kinetic energy of the horizontal and vertical links that
keeps the string oriented in the (1,1) direction.
\vskip -0.5cm

\section{Directed strings and the spin-1 chain} \label{sec_spin1}  
 
Quite generally, the string problem does not simply reduce to that of
the internal dynamics of the worldsheet, because of the requirement 
that the worldsheet has to be embedded in D+1 dimensional space. 
However, in the presence of directedness order and in the absence of 
particle number fluctuations\cite{Karlsruhe} the string boundary
conditions are trivially fulfilled and the string problem is equivalent 
to that of a single ``world sheet `` in 1+1 dimensions. Assume the 
string to be directed along the $x$ direction. Since the string steps 
always forward in this 
direction, the number of particles in the string has to be equal to 
the number of lattice sites in the $x$-direction, and every directed 
string configuration will connect the boundaries in this direction. 
The string is still free to move along the $y$ direction.  Instead of 
labeling the positions in the 2D plane the string is completely 
specified by the list of links, for which there are only 3 
possibilities (in the $(1, 1) $, $(1,0)$, or $(1,-1)$ direction), and the 
position of a  
single ``guider point''.  As a guider point we can take the position
$\bf{r}$ of any one of the particles, which, together with the
relative coordinates given by the links, fixes the position of the
entire string.  Since the guider represents just a single degree of
freedom, and since the thermodynamic behavior of a chain is determined
by the link interactions, the guider coordinates will be irrelevant
for the behavior of the chain.  Apart from this guider degree of freedom
the directed string problem reduces to a one-dimensional quantum
problem with three flavors.

From Eq.\ (\ref{Heff}) one directly deduces the Hamiltonian of the
string directed along $x$,
\begin{eqnarray}
  {\cal H}_{eff} = \sum_{l,k} && \left[ { {\cal K} \over n} \delta(
  |y_{l+1,k}-y_{l,k}| - 1 ) \right.  \nonumber \\ && + {{\cal L}_{12}
    \over n} \delta( | y_{l+1,k} - y_{l-1,k} | - 1 ) \nonumber \\ && +
  {{\cal L}_{22} \over n} \delta( | y_{l+1,k} - y_{l-1,k} | - 2 )
  \nonumber \\ && + \left. \ln ( {{\cal T} \over n} ) \delta ( |
  y_{l,k} - y_{l,k+1} | -1 ) \right].  \label{Hdir}
\end{eqnarray}
It is clear that the directedness simplifies the model considerable.
The directed version can not self-intersect, and the excluded volume
constraint is satisfied automatically.  Furthermore the ${\cal
  L}_{11}$-type of configurations are not allowed, thus the directed
model is specified by three parameters and the temperature (${\cal T}
= 1$). Because of the preceding considerations, Eq. (\ref{Hdir})
corresponds with a 1+1 D problem, which is actually equivalent to a
general quantum spin-1 chain.

\begin{figure}

%\vspace{0.5ex}
\epsfig{figure=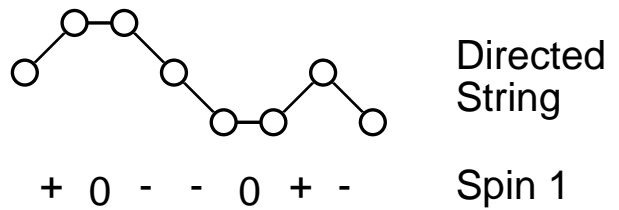,width=8cm}   
%\vspace{1.5ex}
\caption{ The relation between spin-1 and directed strings, $S^z_l =
y_{l+1}-y_l$. }
\label{fig_spinstring}
\end{figure}

We identify the spin with the string {\it height difference}
$y_{l+1}-y_l$, which can be either $0$, $1$ or $-1$, see Fig.\ 
\ref{fig_spinstring}.
  These link dynamical variables specifying the
string can be directly identified with the $m_s = 0, \pm 1$ variables
of the spins living on the sites of the spin chain. Defining the
latter using hard core bosons $b^{\dagger}_{m_s}$, the spin operators
for the $S=1$ case become $S^z = b^{\dagger}_1 b_1 - b^{\dagger}_{-1}
b_{-1}, S^{+} = \sqrt{2} ( b^{\dagger}_1 b_0 + b^{\dagger}_0 b_{-1}) $
and by comparing the action of the spin- and string operators on their
respective Hilbert spaces one arrives at operator
identities\cite{nijsrom}.  A quantum hop from $y$ to $y+1$ increases
the height difference on the left of $l$ by one, and decreases it by
one on the right, as is easily seen by inspecting the two hopping
terms in Fig.\ \ref{fig_energies}.  Therefore\cite{nijsrom},
\begin{eqnarray}
  S^z_l &=& y_{l+1} - y_{l} \nonumber \\ S^{\pm}_{l-1} S^{\mp}_{l} & =
  & 2 P_{Str}(y_l-y_{l-1}) P_{Str}(y_{l+1}-y_{l}) e^{ \pm i{\hat
      \pi}_l}.
                                                         \label{spindef}
\end{eqnarray}
The following identities, for $S=1$,
\begin{eqnarray}
  \delta ( |y_{l+1}-y_l| - 1 ) & = & ( S^z_l )^2 ~,  \nonumber \\ \delta (
  |y_{l+1}-y_{l-1}| - 1 ) & = & (S^z_l)^2 + (S^z_{l-1})^2 - 2 (S^z_l
  S^z_{l-1})^2~, \nonumber \\ \delta ( |y_{l+1}-y_{l-1}| - 2 ) & = & {1
    \over 2} S^z_l S^z_{l-1} [ 1 + S^z_l S^z_{l+1} ]~,
  \label{identities}
\end{eqnarray}
are easily checked.  The directed string problem can now be
reformulated in spin language as,
\begin{eqnarray}
  {\cal H}_{spin}  = & \sum_l   &\left[ ( {\cal K} + 2 {\cal L}_{12} ) (
  S^z_l)^2 + { { {\cal L}_{22}} \over 2 } S^z_l S^z_{l-1} \right. 
  \nonumber \\ &  + & \left( { {{\cal L}_{22}} \over 2} - 2 {\cal
    L}_{12} \right) ( 
  S^z_l S^z_{l-1} )^2  \nonumber \\
  &  + & \left. { {\cal T} \over {2}} (S^+_l S^-_{l-1}+
  S^-_{l} S^+_{l-1})  \right]~. \label{Hspin}
\end{eqnarray}
Following the spin-1 literature we define the following parameters,
\begin{eqnarray}
  D &=& {\cal K} + 2 {\cal L}_{12}~, \nonumber\\ J &=& { {{\cal L}_{12}}
    / 2}~, \nonumber\\ E &=& {{\cal L}_{22}} / 2 - 2 {\cal L}_{12}~.
  \label{DJE}
\end{eqnarray}
The $E$ term is new.  It is a quartic Ising term, leading to extra
phases and phase transitions.  For the special choice $E = 0$ (${\cal
  T}=1$), the above Hamiltonian reduces to the familiar XXZ model with
on-site anisotropy,
\begin{eqnarray}
  {\cal H}_{XXZ} = && \sum_l \left[ D (S^z_l)^2 + J S^z_l S^z_{l-1}
\right. \nonumber \\ && +{1 \over 2} \left. (S^+_l S^-_{l-1} + S^-_{l}
S^+_{l-1}) \right].  \label{XXZ}
\end{eqnarray}
The zero temperature phase diagram of the above spin-1 model has been
discussed in detail in the literature.
\cite{Bot83,Gla84,Sch86,Sch86b,nijsrom} In the next section we will
briefly review the six phases found for this model, from a string
perspective.  Then we will show that a nonzero $E$ parameter leads to
the appearance of four extra phases in section VII.

Den Nijs and Rommelse \cite{nijsrom} discuss a direct mapping between
the spin chain and the RSOS surface.  We stress that this mapping in
fact involves two steps: First the RSOS model is mapped on a string
problem, using the $T$ matrix.  Then the spins are identified as shown
above.  Thus the quantum string is a {\it natural intermediate} of the
two other models.  Den Nijs and Rommelse make use of the freedom in
the choice of the $T$ matrix to define a mapping which is slightly
different from ours, since they introduce a transfer matrix along a
diagonal, while we introduce one along the $x$-direction. As a result, 
in their case there are only interactions between  next nearest neighbors 
along the $(1,1)$ direction,  while our choice allows for interactions 
between next nearest neighbors along the $x$-direction. Therefore, our 
RSOS model differs slightly from theirs.

The RSOS representation is more transparent than the quantum model.
The spin-1 phases and the nature of the phase transitions all have a
natural interpretation in space-time.  For instance the Haldane phase,
or AKLT wavefunction, with its mysterious hidden string order
parameter is identified as a ``disordered flat'' RSOS surface
\cite{nijsrom} with a simple local order parameter.  The height
representation, dual to the spins, gives a similar local order
parameter for the quantum string.

\section{The phases (E=0)}\label{sec_phases} 

In this section the general string Hamiltonian will be simplified by
leaving out the quartic Ising term ($E=0$ in Eqs.\ (\ref{Hspin}) and
(\ref{DJE})).  Our string problem is now equal to the spin-1 XXZ
model.  The zero-temperature phase diagram of the string problem is
surprisingly rich, and even for the case $E=0$ there are 6 phases and
a large variety of phase transitions.  These phases can be classified
in three groups: classical strings localized in space, quantum rough
strings of the free variety, and partly delocalized phases of which
the disordered flat phase is a remarkable example.  In this section we
will briefly review the six phases as discussed in the literature on
the spin-1 XXZ problem \ (\ref{XXZ}).  The problem will be addressed
from the quantum string perspective.  For more details we refer to
Ref.\ \onlinecite{nijsrom}.  In the next section we will show that
with a finite $E>0$ four additional phases are stabilized. 

\begin{figure}
%\vspace{0.5ex}
\epsfig{figure=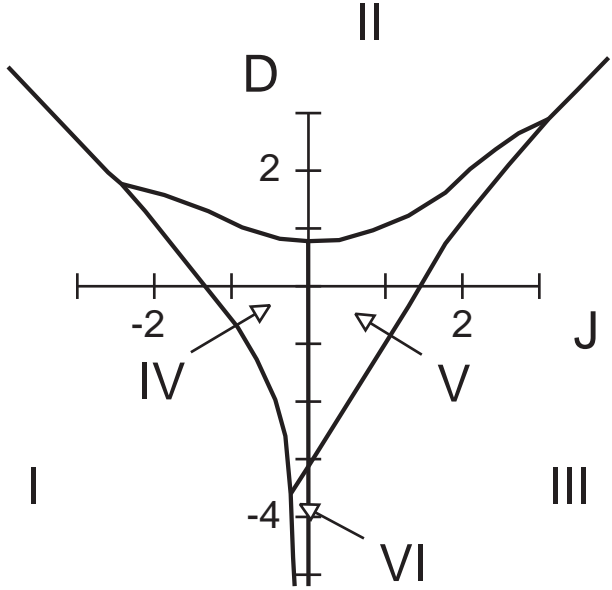,width=\linewidth}   
%\vspace{1.5ex}
\caption{ The phases and phase transitions of the directed quantum
  string as a function of the  on-site anisotropy $D$ and the Ising
  interaction $J$ of the corresponding spin-1 XXZ model. The parameter 
$E$ is set to zero. }
\label{fig_pd0}
\end{figure}

The phase diagram of the quantum string is shown in Fig.\ \ref{fig_pd0}, 
as a function of $D$ and $J$.  We have used the XXZ
model parameters, defined in Eq.\ (\ref{DJE}), such that the phase
diagram can be compared directly with the spin-1 literature
\cite{Bot83,Gla84,Sch86,Sch86b} and in particular with Fig.\ 13 of
Ref.\ \onlinecite{nijsrom}. We will introduce below the various order
parameters that have been introduced in this reference to distinguish
the six phases in this phase diagram. The relation between the more
general ($E\neq 0$) string and spin phases will be clarified in the
next section.

There is first of all a horizontal and a diagonal string phase.  In
the diagonal phase I no quantum fluctuations are allowed, since, as
shown in Fig.\ \ref{fig_Tblocks} a diagonal string does not couple to
other states by ${\cal H}_Q$.
 This phase is stabilized by a large and
negative ${\cal L}_{22}$, so that since $E=0$ also $J={\cal L}_{12}/2
={\cal L}_{22}/8$ is large and negative. A suitable variable 
introduced to define order parameters, following Ref. \onlinecite{nijsrom},  
is the Ising spin variable $\sigma_l= (-1)^{y_l}$, which identifies whether 
a given height is in an even or odd layer. This underlying spin model
can have ``ferromagnetic'' or antiferromagnetic'' order, and so we
introduce the corresponding order parameters\cite{noteheight}
\begin{eqnarray}
  \rho & =& \langle \sigma_l \rangle~, \hspace*{6mm} \rho_{stag}= \langle (-1)^l
  \sigma_l \rangle~ ,\nonumber \\ & & \hspace*{4mm} \rho_{str}   =  \langle \sigma_l
  (y_{l+1}-y_l)\rangle~.\label{rhos}
\end{eqnarray} 
\noindent Here the brackets denote the ground state expectation value as well as
an average over string members $l$. In (\ref{rhos}), we have also
included the order parameter $\rho_{str}$ discussed below.  In the
horizontal phase II one particular height is favoured, thus the order
parameter $\rho$ is nonzero here.  This phase is stabilized by a large
positive ${\cal K}$, which suppresses diagonal links.  However ${\cal
  H}_Q$ causes virtual transitions from two horizontal links into two
diagonal ones, see Fig.\ \ref{fig_energies}.  On the 2D worldsheet
these fluctuations show up as local terraces that do not overlap and
thus do not destroy the long-range order.  In both phases the
elementary excitations are gapped.

Upon lowering ${\cal K}$ the terraces grow and at some point they will
form a percolated network --- the string has become disordered in both
space and imaginary time.  Via the well-known Kosterlitz-Thouless
roughening transition\cite{Wee80} phase IV is entered for $J<0$.  This
phase belongs to the well known XY universality class, characterized
by algebraic correlation functions and gapless meandering excitations
--- capillary waves in fluid interface language.  The roughness,
however, is extremely ``soft'' and the height difference diverges only
logarithmicly, $\langle (y_l-y_m)^2\rangle \sim \ln |l-m|$.  The
transition from the Gaussian phase which is rough and on average
oriented horizontally to the ``frozen'' diagonal phase is a ``quasi
first-order'' KDP transition\cite{nijsrom}.

For large negative ${\cal K}$ diagonal links are favored over
horizontal ones.  There is a transition to a second rough phase (phase
VI).  It is distinguished from the first by the order parameter
$\rho_{stag}$, which is zero in phase IV.  In this phase horizontal
links are virtual and occur in pairs. As we will discuss later in
section \ref{sec_pd}, for large negative $K$ the model can therefore
be reduced to an effective spin 1/2 problem.

For negative ${\cal K}$ and positive $J$ (=${\cal L}_{12}/2 = {\cal
  L}_{22}/8$) the string becomes a (physically unlikely) zigzag with
alternating up and down diagonal pieces.  Excitations to pairs of
horizontal links are gapped.  Again $\rho_{stag} = \langle
(-1)^{-l}\sigma_l \rangle$ serves as an order parameter. Upon increasing
${\cal K}$ the islands formed by pairs of horizontal links start to
overlap and there is an Ising transition into the Haldane or
disordered flat (DOF) phase.

The point $J=1, D=0$ belongs to the gapped DOF phase, in agreement
with Haldane's educated guess\cite{Hal83,affl89} that integer spin
chains are gapped at the Heisenberg antiferromagnetic point.  In this
``disordered horizontal'' string phase the prototype wavefunction,
equal to the AKLT valence bond state\cite{Aff87}, has every up
diagonal link followed by a down link, with a random number of
horizontal links in between.  The height $y_l$ takes just two values,
say 0 and 1.  The local order parameter is, $\rho_{str}$, defined in
(\ref{rhos}). This order parameter measures the correlation between the
next step direction and whether one is in a layer of even or odd
height. When $\rho_{str}=1$, the string just steps up and down between
two layers, but the steps can occur at arbitrary positions. Note that
the height is a global quantity in spin language, i.e., it is the
accumulated sum over spins, $y_l = \sum_{m=0}^l S^z_m$.  Because of
this the above order parameter becomes non-local when rewritten in
terms of the ``string'' of spins. Therefore, it is often called the
string order parameter. We will also use this name, but stress that
the ``string of spins'' to which this name refers should not be
confused with the general strings which are the basis of our model,
and that the other order parameters are nonlocal as well in terms of
the original spins $S$.

\begin{figure}

%\vspace{0.5ex}
\epsfig{figure=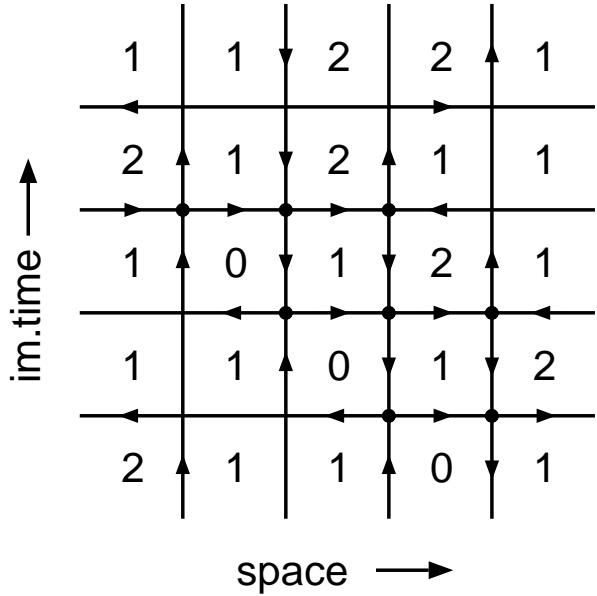,width=8cm}   
%\vspace{1.5ex}
\caption{ Vertices (thick dots) on the space, imaginary time string
worldsheet.  The numbers correspond to the heights $y_{l,k}$.
Arrows are drawn when the heights of neighbors differ.  When four
arrows occur at a crossing point this is called a vertex.}
\label{fig_vertex}
\end{figure}

This phase diagram can be rationalized by writing the RSOS problem as
the product of a 6-vertex model and the 2D Ising model of $s$ spins on
the 6-vertex lattice, as discussed in detail by Den Nijs and Rommelse.
\ \onlinecite{nijsrom}.  The horizontal, diagonal, zigzag but also
the second rough phase VI all correspond to Ising order: $\rho =
\langle \sigma_l \rangle $ is nonzero in the horizontal phase II,
while $\rho_{stag}= \langle (-1)^{l}\sigma_l \rangle$ is nonzero in
the diagonal phase I, the zigzag phase III and the rough phase VI.
The six-vertex part is defined on the crossing points of steps on the
surface --- see Fig.\ \ref{fig_vertex}.
 This is a (sometimes highly)
diluted set of points.  The Ising degree of freedom disorderes on the
transition between the phases III and V, and between IV and VI, while
the six-vertex part remains unchanged.  Therefore these transitions
are Ising like.  Transitions I $\rightarrow$ IV, I $\rightarrow$ VI ,
IV $\rightarrow$ V and III $\rightarrow$ VI are related to the
six-vertex part becoming critical, and these KDP and KT transitions
are known from the quantum spin-1/2 chain.  The transition II to IV is
related to the famous surface-roughening transition, of the
Kosterlitz-Thouless type.\cite{Wee80,Bei87} The subtle transition between
phase II and V, is coined ``preroughening transition'' by den Nijs. 
It separates two gapped phases.  At the transition the gap closes and
the system is Gaussian, with varying exponents along the transition
line.\cite{Bot83,Gla84,Sch86,Sch86b}
\vskip -0.3cm

Almost all the phases can be distinguished by the above order
parameters $\rho$, $\rho_{stag}$ and $\rho_{str}$, except that these
do not discriminate between the diagonal phase I and the rough phase
VI. These two phases can be identified by also introducing an order
parameter which detects the presence of an average slope,
$\rho_{slope}= \langle y_{l+1 } -y_l\rangle $. In Table
\ref{tabE0Phases} we list the various phases for $E=0$ and the order
parameters.
 
As we shall see in the next section, in the general case $E\neq0$ it
is more convenient to introduce slightly different spin variables to
identify all the ten different phases that occur then. The choice of
Ref. \onlinecite{nijsrom} discussed here is somewhat more convenient
for understanding the universality classes of the various phase
transitions.

\section{The full phase diagram -- phase-boundary estimates}\label{sec_pd}

As mentioned above the quartic Ising term with prefactor $E$
generalizes the XXZ Hamiltonian and leads to extra phases.  We will
motivate that four extra phases are to be expected, and we will show
that they are stabilized by a positive $E$ parameter.  The most
disordered phase is still the Gaussian phase.

\begin{figure}
%\vspace{0.5ex}
\epsfig{figure=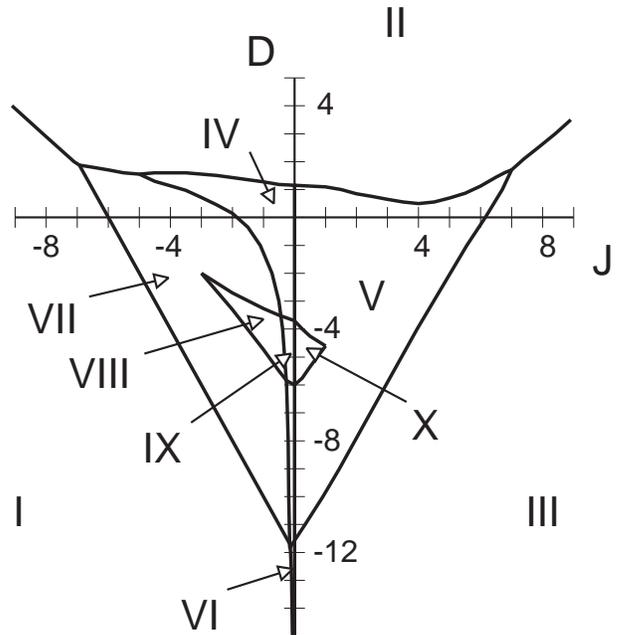,width=\linewidth}   
%\vspace{1.5ex}
\caption{ The phases and phase transitions of the quantum string for
$E=5$ as a function of t.  On the axis are the on-site anisotropy 
$D$ and the Ising interaction $J$.  }
\label{fig_pd5}
\end{figure}

Using a similar decomposition as above, we can determine how many
different phases to expect for a general spin-1 chain with $z$-axis
anisotropy and nearest-neighbor interactions\cite{Sch95}. Think of the
spin 1 as consisting of two spins ${1\over 2}$, see Table\ \ref{tabSpinhalf}.
The first is $\sigma^z=\downarrow$ when the spin 1 has $S^z=0$ and
$\sigma^z=\uparrow$ when $S^z=\pm 1$, similar to the Ising degree of
freedom defined above. This spin thus indicates the presence or
absence of a step.  The second spin ${1\over 2}$ ${\bf s}$, is defined 
as $s^z = S^z/2$ when $S^z=\pm 1$ and is absent when $S^z=0$.  This is 
related to the diluted vertex network discussed by den Nijs and Rommelse, 
in that if there is a step, the $z$-component of ${\bf s}$ indicates
wether this step is up or down.  The spins ${\bf s}$ can have
ferromagnetic (F) or antiferromagnetic (AF) order, or they can be
disordered (D).  For ${\bf \sigma}$ the two ferromagnetic cases
correspond to different physical situations, and we have to
distinguish ferromagnetic $\downarrow$ (F1), a horizontal string, from
ferromagnetic $\uparrow$ (F2).  When ${\bf \sigma}$ has F1 order,
${\bf s}$ becomes irrelevant (or better --- there are disconnected
finite terraces of ${\bf s}$ spins with short-range correlations).
Therefore one expects $10$ phases, depending on the order of the two
spin species: 1 F1 phase, 3 F2 phases, 3 $\sigma$-disordered phases,
and 3 $\sigma$-antiferromagnetically ordered phases.  These are listed
in Table\ \ref{tabPhases}. An example of a phase diagram in a case in
which all ten phases are present is show in Fig.\ \ref{fig_pd5}, which 
corresponds to the case $E=5$. The detailed of how this phase  
diagram was obtained will be discussed below. 
\vskip -0.5cm

There are four new phases, VII to X, compared to the phase diagram
discussed in the previous section.  All four are stabilized by a
positive $E$ parameter in Eq.\ (\ref{Hspin}).  Three phases , VIII to
X, result from an antiferromagnetic order of the ${\bf \sigma}$ spin.
This corresponds to alternating horizontal and diagonal string links
(see Table \ \ref{tabPhases}).  The diagonal links can be either all
up (FM, phase VIII), alternatingly up and down (AFM, phase X) or
disordered (phase IX).  In phase VII the ${\bf \sigma}$ spin is
disordered, while the ${\bf s}$ spin is in the FM phase.  This is a
diagonal wall diluted with horizontal links.  These links coherently
move up and down along the wall, lowering the kinetic energy.  The
wall can take any average angle between $-\pi/4$ and $\pi/4$, and this
angle is fixed by the value of the parameters.  We will call this the
``slanted'' phase.  In terms of the decomposition into an Ising spin
model and a six-vertex model of den Nijs and Rommelse it is easy to
see that the horizontal links change the orientation of the Ising spin
and act like a Bloch wall.  The Ising spin is therefore disordered.
The six-vertex term is irrelevant for the existence of the slanted
phase --- in the case of a single horizontal link, i.e., on the
boundary between the slanted and diagonal string phase, there are no
vertices.

A large part of the phase boundaries can be estimated exactly, almost
exactly, or to a fair approximation.  Let us focus first on the
classical phases.  The diagonal, horizontal and zigzag phases have the
following energies in the classical approximation in which there are
no fluctuations, as is easily verified,
\begin{eqnarray}
  {E}_{I} &=& L( {\cal K} + {\cal L}_{22} ) = L( D + J + E )~,
  \nonumber
  \\ {E}_{II} &\approx& 0~, \nonumber \\ {E}_{III} &\approx& L {\cal K}
  = L( D - J + E )~.  \label{E123}
\end{eqnarray}
where $L$ is the length of the chain.  The first-order transitions
will therefore occur close to the lines ${\cal K} = -{\cal L}_{22}$
($D = -J-E$) between phases I and II, ${\cal L}_{22} = 0$ ($J = 0$)
between phases I and III, and ${\cal K} = 0$ ($D = J-E$) between
phases II and III. These transitions become exact in the classical or
large-spin limit.

The transition between phase I and VII, the diagonal and slanted
phase, can be found exactly.  The transition is of the
Pokrovsky-Talapov, or conventional 1D metal-insulator type (see, for
instance, Ref.\ \onlinecite{dennijsrev}).  The horizontal link can be
seen as a hard-core particle or a spinless fermion, with the parameters
determining an effective chemical potential.  For a critical chemical potential
equal to the bottom of the band of the hard-core particle the band
will start to fill up.  The transition occurs when the diagonal string
becomes unstable with respect to a diagonal string with one horizontal
link added.  This single link delocalizes along the string with a
momentum $k$ and a kinetic energy $2{\cal T} \cos(k)$.  The minimal
energy is $(L-1){\cal K} + (L-2){\cal L}_{22} + 2{\cal L}_{12} -
2{\cal T}$, and the transition occurs when ${\cal K} = 2({\cal L}_{12}
- {\cal L}_{22} - {\cal T})$ or, with ${\cal T}=1$, when
\begin{equation}
  \mbox{I to VII transition:} \hspace*{4mm} D = -2(J+E+1)~.
  \label{ItoVII}
\end{equation}

The transition between phase III and V will occur when horizontal link
pairs unbind in the zigzag background.  A rough estimate, neglecting
fluctuations, is obtained by comparing the energy of a single
horizontal link with that of a perfect zigzag.  In the same way as
above we estimate the phase boundary to be close to $ D = 2(J-E-1)$.
In the same way the transition from phase II to V or IV is determined
by the energy of a single diagonal step in a horizontal wall, which
becomes favourable when $D = 2$.  This last estimate turns out to be
very crude, in that it largely underestimates the stability of the
flat phase.

For large negative ${\cal K}$ the horizontal links are strongly
suppressed, and the string can be mapped perturbatively on a spin 1/2
chain.  Identify $S^z=1$ (diagonal upward) with $s^z=\uparrow$ and
$S^z=-1$ (diagonal downward) with $s^z=\downarrow$.  Via a virtual
(0,0) spin pair (two horizontal links) the spins can still fluctuate,
(1,-1) $\rightarrow$ (0,0) $\rightarrow$ (-1,1).  One finds, using
second-order perturbation theory in ${\cal T}/{\cal K}$,
\begin{eqnarray}
  {\cal H}_{eff}(D \rightarrow -\infty) &=& (4J+j_{\pm}) \sum\limits_l
  s^z_l s^z_{l+1} \nonumber \\ && + j_{\pm} {1 \over 2} \sum\limits_l
  \left( s^+_l s^-_{l+1} + s^-_l s^+_{l+1} \right)~,\nonumber \\ 
    j_{\pm} & = & \frac{2{\cal T}^2}{|2D+3E|}~. \label{HKneg}
\end{eqnarray}
Here we subtracted an irrelevant constant term.  This has the form of
the well studied spin 1/2 Heisenberg chain with Ising anisotropy.
Transitions occur when $(4J+j_{\pm}) = \pm j_{\pm}$, or when $J=0$
(III to VI) and $J=-1/|2D+3E|$ (I to VI) (setting ${\cal T}=1$).

The above estimates seem to suggest that the line $J=0$ is special.  
Our numerical results show that it describes accurately the transition
between III and VI, but also the transition between IV and V. This
agrees with the arguments given by Den Nijs and Rommelse\cite{nijsrom}
that the Kosterlitz-Thouless transition between IV and V should occur
precisely at the $J=0$ line.

The slanted phase consists predominantly of up diagonal and horizontal
links.  Neglecting down diagonals altogether, which turns out to be a
good approximation, one can again map the string or spin-1 chain on an
effective spin 1/2 system.  Now the relevant degree of freedom is the
$\sigma$ Ising degree of freedom.  Because $\sigma = \uparrow$ (a
diagonal link) is not symmetrically equivalent to $\sigma =
\downarrow$ (a horizontal link) the spins will feel an effective
magnetic field, which regulates the density of horizontal links.
Rewriting Eq.\ (\ref{Hspin}) gives,
\begin{eqnarray}
  {\cal H}_{eff} &=& D \sum\limits_l (\sigma^z_l + {1 \over 2})
  \nonumber \\ && + (J+E) \sum\limits_l (\sigma^z_l + {1 \over 2})
  (\sigma^z_{l+1} + {1 \over 2}) \nonumber \\ && + {\cal T}
  \sum\limits_l \left( \sigma^+_l \sigma^-_{l+1} + \sigma^-_l
  \sigma^+_{l+1} , \right)~, \label{HJMcC}
\end{eqnarray}
and, after rescaling and putting ${\cal T}=1$,
\begin{eqnarray}
  {\cal H}_{eff} &=& h \sum\limits_l \sigma^z_l + \Delta \sum\limits_l
  \sigma^z_l \sigma^z_{l+1} \nonumber \\ && + {1 \over 2}
  \sum\limits_l \left( \sigma^+_l \sigma^-_{l+1} + \sigma^-_l
  \sigma^+_{l+1} \right)~, \label{HJMcC2}
\end{eqnarray}
with the field $h = (D+J+E)/2$ and Ising coupling $\Delta = (J+E)/2$.
On the line $h=0$ the number of up diagonal links equals the number of
horizontal links. The average tilt angle is thus 22.5$^o$ in this
approximation.  The phase diagram of the spin 1/2 chain in the
$h$-$\Delta$ plane was discussed by Johnson and McCoy\cite{Joh72}.
For $h=0$ there are three phases.  The ferromagnet corresponds to
phase I, the antiferromagnet with phase VIII, and the gapless
disordered phase translates to the slanted string phase VII.
Increasing the field $h$ in the AFM phase will cause a transition to
the gapless phase with a finite magnetization.  In the approximation
that down diagonals are neglected, it follows from the results of
Johnson and McCoy\cite{Joh72} that the point $\Delta=1, h=0$ or
$J=2-E$ is the point with the most negative value of $J$ where phase
VIII is stable.  For $E=0$ (as well as for small values of $E$) this
occurs in the positive $J$ side of the phase diagram, meaning that
phase VIII to X will in fact not be stable: for positive values of
$J$, down steps in the original model proliferate.  To have a phase
diagram with all 10 phases present we choose $E=5$.

\vskip -0.5cm
\begin{figure}

%\vspace{0.5ex}
\hspace{-2ex}\epsfig{figure=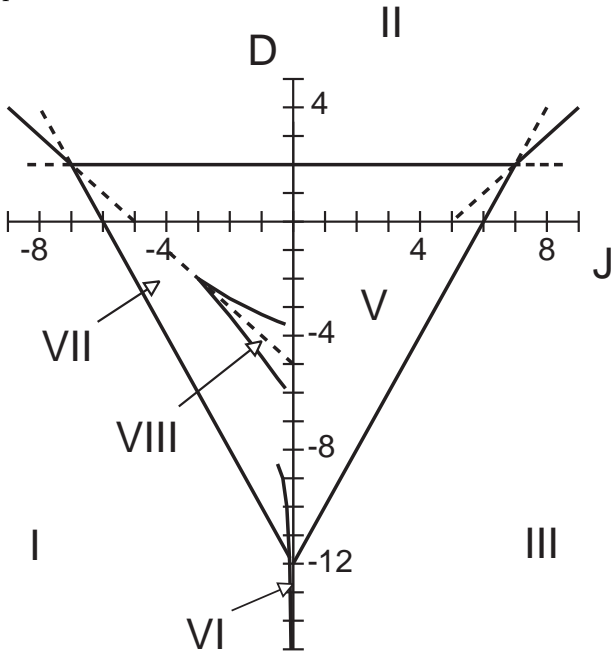,width=\linewidth}   
%\vspace{1.5ex}
\caption{ The various phase transitions, obtained from semiclassical
estimates, exact arguments and perturbative mappings to spin 1/2.}
\label{fig_estimates}
\end{figure}
In Fig.\ \ref{fig_estimates} the various phase-boundary estimates
given above are summarized.  The topology of the main part of the
phase-diagram has now become clear.  In the centre of the figure for
$E=5$ the Johnson-McCoy phase diagram is inserted.  The estimates
suggest that at least phase VII and VIII are stabilized by taking
$E=5$.  The dotted line through phase VIII is the line where the
effective field $h$ is zero, and the number of diagonal links is
(nearly) equal to the number of horizontal ones.

We finally argue that the slanted phase exists in some region of the
phase diagram for any $E>0$. To see this, first consider the case
$E=0$. Along the line $D=0$, our model (with $E=0$) corresponds to the
Heisenberg model with Ising anisotropy, and the point $J=-1$
corresponds to the isotropic ferromagnetic Heisenberg point. Along the
line $D=0$, the transition from the ``ferromagnetic'' (diagonal in
string language) phase I to the XY phase IV therefore occurs at
$J=-1$. Now, for $E=0$, the exact location of the line along which
phase I becomes unstable to the slanted phase VII, given by
$D=-2(J+1)$ according to (\ref{ItoVII}), goes exactly through the
ferromagnetic Heisenberg point at $J=-1$ as well. The results of Fig.
13 of den Nijs and Rommelse\cite{nijsrom} indicate that this line then
touches the phase boundary between phases I and IV right at this point
in such a way that for $E=0$ no slanted phase occurs. If we assume
that both phase boundaries shift linear in $E$ for $E$ nonzero and
small, it is clear that the slanted phase must stabilize in some
region near the point $D=0, J=-1$ for one sign of $E$, while for the
other sign the phase must be absent. Physically, it is clear that the
stabilization of the slanted phase will occur for positive values of
$E$, $E>0$.

\section{Numerical}\label{sec_num}

To fill in the details of the phase diagram we have performed exact
diagonalization and finite-temperature quantum-Monte Carlo
calculations.  Ground state properties of strings up to 15 holes (spin
chains of length 14) were obtained using the Lanczos diagonalization
method.  For the Monte-Carlo we used the checkerboard decomposition,
briefly explained in Sec.\ \ref{sec_RSOS}.  The Monte Carlo has the
disadvantage that an extra limit to zero temperature has to be taken,
a regime where the updating slows down considerably and where it is
difficult to judge the accuracy.  For determining the phase boundaries
of the directed string we mainly used the Lanczos results for the
equivalent spin model.  On the
other hand the Monte Carlo space-time worldsheets provide a
transparent physical insight into the phases, phase transitions, order
parameters, etcetera. Moreover, the Monte Carlo method allows, of
course, to
 treat bigger systems.

We are in the fortunate situation that the order parameters of the
various phases and the universality classes of the transitions are
known.  This offers a variety of approaches to determine the critical
lines --- one can monitor the finite-size behavior of the order
parameter, correlation functions or the energy level spacings.
Typically we applied two independent methods to the various
transitions.  Our aim is to map out the entire phase diagram with an
accuracy of roughly the line thickness in the phase diagram.  For very
accurate estimates other methods, notably the density-matrix
renormalization group treatment of White\cite{Whi92}, are more
appropriate.

An elegant and powerful method is the
phenomenological renormalization group approach pioneered by
Nightingale.\cite{Nig82,Bar83} In this approach one consideres an
infinite strip with a width $L$, as a finite-size approximation to
the 2D classical system.  At the critical temperature of the
infinite system one expects, from finite size scaling, that
the correlation length along the strip scales like the width of the
strip ($L_1$ or $L_2$),
\begin{equation}
    \xi_{L_1}(T_c) = \frac{L_1}{L_2} \xi_{L_2}(T_c).
        \label{frg}
\end{equation}
The infinite strip is solved by diagonalizing the (finite) $T$ matrix.
The correlation length can be calculated from,
\begin{equation}
        \xi = 1 / \ln(\lambda_1/\lambda_0),
        \label{xi}
\end{equation}
where $\lambda_0$ and $\lambda_1$ are the largest and second largest
eigenvalues of the $T$ matrix.

\begin{figure}
%\vspace{0.5ex}
%\hspace{-2ex}\epsfig{figure=fig_frg.eps,width=\linewidth}   
\epsfig{figure=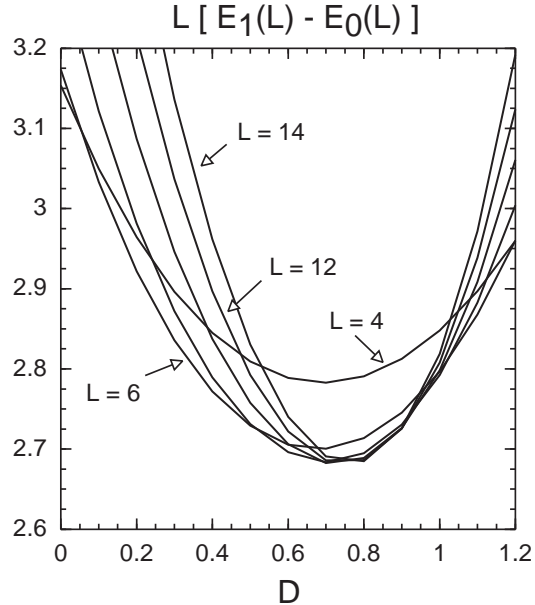,width=0.8\hsize}   
%\vspace{1.5ex}
\caption{ Estimate of the preroughening transition between phase II
and V. The plot shows $L [ E_{1}(L) - E_{0}(L) ]$ for various lengths
$L$, as a function of the parameter $D$, with $J= 0.8$ and $E=0$.  The
two crossing points between the successive curves form an upper and
lower estimate of the transition. }
\label{fig_frg}
\end{figure}

A finite 1D quantum chain can be viewed as a strip infinitely long in
the imaginary time direction (zero temperature).  In the time
continuum limit, writing $T = \exp(\tau H)$, the equation
corresponding to Eqs.\ (\ref{frg}) and (\ref{xi}) is,
\begin{equation}
    L_1 [ E_{1}(L_1) -  E_{0}(L_1) ] = L_2 [ E_{1}(L_2) -  E_{0}(L_2) ],
        \label{frgQ}
\end{equation}
for two different string lengths $L_1,L_2 \rightarrow \infty$ and for
parameter values at criticality.  Here $E_0$ and $E_1$ are the ground
state and first excited energy of a quantum Hamiltonian $H$. According
to (\ref{frgQ}), a phase transition line can be located by studying
the energy gap as a function of $L$, while  monitoring when
(\ref{frgQ}) is obeyed. A practical
example of Eq.\ (\ref{frgQ}) is shown in Fig.\ \ref{fig_frg}.
Successive curves for lengths 
$L$ and $L+2$ show two crossing points.  Extrapolating these crossing
points to infinite $L$ gives two estimates of the
preroughening transition from phase II to V. 

The above scaling holds when time and space scale in the same way, or $
(E_{1}(q) - E_{0}) \sim q^z$, with a dynamical exponent $z=1$.  This
can be checked independently, giving a self-consistent justification
of the use of Eq.\ (\ref{frgQ}).  In 
a similar spirit one can determine the critical scaling of correlation
functions of an operator $O$ by monitoring the lowest energy state
with a nonzero overlap with $O|0\rangle$.  This has been used
extensively\cite{Bot83,Gla84,Sch86,Sch86b} to study the phases and
exponents of the Gaussian phases of the spin-1 chain.  We refer to
these articles for more details.  

Another method used to determine second-order phase transitions is
the Binder parameter\cite{Bin81}, which we define as $ (3-\langle m^4
\rangle / \langle m^2 \rangle^2)/2$, where $m$ is the relevant order
parameter.  This quantity tends to 1 in the ordered phase, where
$\langle m^4 \rangle \approx \langle m \rangle^4$, and approaches 0 in
the disordered phase, where $m$ has a Gaussian distribution around
$\langle m \rangle = 0$.  In a renormalization group sence the shape
of the order parameter distribution function becomes independent of
the size at criticality.  The various curves for different sizes of
the Binder parameter versus model parameters or $T$ should therefore
cross at the critical point.  For instance for the Ising transition
between phase III and V we take $m = (-1)^{y_l-l}$, the Ising degree
of freedom introduced before.
\vskip -0.3cm

\begin{figure}

%\vspace{0.5ex}
%\hspace{-82ex}\epsfig{figure=fig_KT.eps,width=\linewidth}   
\hspace{-72ex}\epsfig{figure=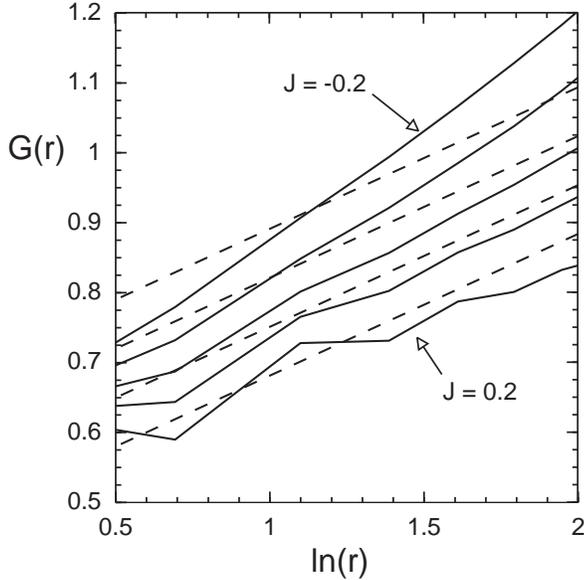,width=0.9\linewidth}   
%\vspace{1.5ex}
\caption{ Estimate of the Kosterlitz-Thouless transition between phase
IV and V. At the transition the slope between $G(r) = \langle
(y_r-y_0)^2 \rangle $ and $\ln(r)$ approaches $2/\pi^2$ (the
dotted lines).  }
\label{fig_KT}
\end{figure}

The Kosterlitz-Thouless transition between for instance IV and V is a
subtle one due to the infinite order of the transition and the
exponential vanishing of the gap.  Previous studies show a large
uncertainty in the position of the transition line.  In the entire
Gaussian rough string phase the system is critical.  Height
correlations diverge very weakly like $G(r) = \langle (y_r-y_0)^2 \rangle
\propto C \ln(r)$.  At the KT transition point the prefactor takes the
universal\cite{Bei87} value $C = 2 / \pi^2$.  We found that this
relation is very useful to determine the KT transition line --- see 
Fig.\ \ref{fig_KT}.
  This relation gives surprisingly good
results even for very small distances, and is consistent with a KT
transition at $J=0$, as discussed by den Nijs and Rommelse.\cite{nijsrom}

The phase diagram for $E=5$ is shown in Fig.\ \ref{fig_pd5}.  For this
value of $E$ the slanted phase VII is very pronounced.  The phases
VIII to X occur in a small region in the middle of the diagram around
the line of equal probability of horizontal and diagonal links, in the
approximation that the Hamiltonian can be mapped onto the spin 1/2
problem (\ref{HJMcC2}).
The rough phases occur at small negative values of $J$.  Phases I, II,
III, V, VIII and X are gapped.
\vskip -0.5cm

The character of the various phases becomes clear by looking at the
corresponding Monte-Carlo snapshots in Fig.\ \ref{fig_snap}.

 In the flat phase one particular height $y_l$ dominates, and the quantum
fluctuations do not percolate in space-time.  Decreasing ${\cal K}$ 
these quantum fluctuation islands grow and when they overlap the
string enters phase IV. The system is Gaussian rough in both space and 
time. Note the very weak logarithmic meandering --- e.g. for a string 
of length 10, Fig.\ \ref{fig_KT} shows that the mean square height 
fluctuations are only of order 1 near the KT transition. Despite this, 
determining  the KT transition from $G(r)$ works surprisingly well for 
the small  systems calculated.

%\vskip -4.6cm
(a)
\vskip -0.5cm
\begin{figure}

\hspace{-2ex}\epsfig{figure=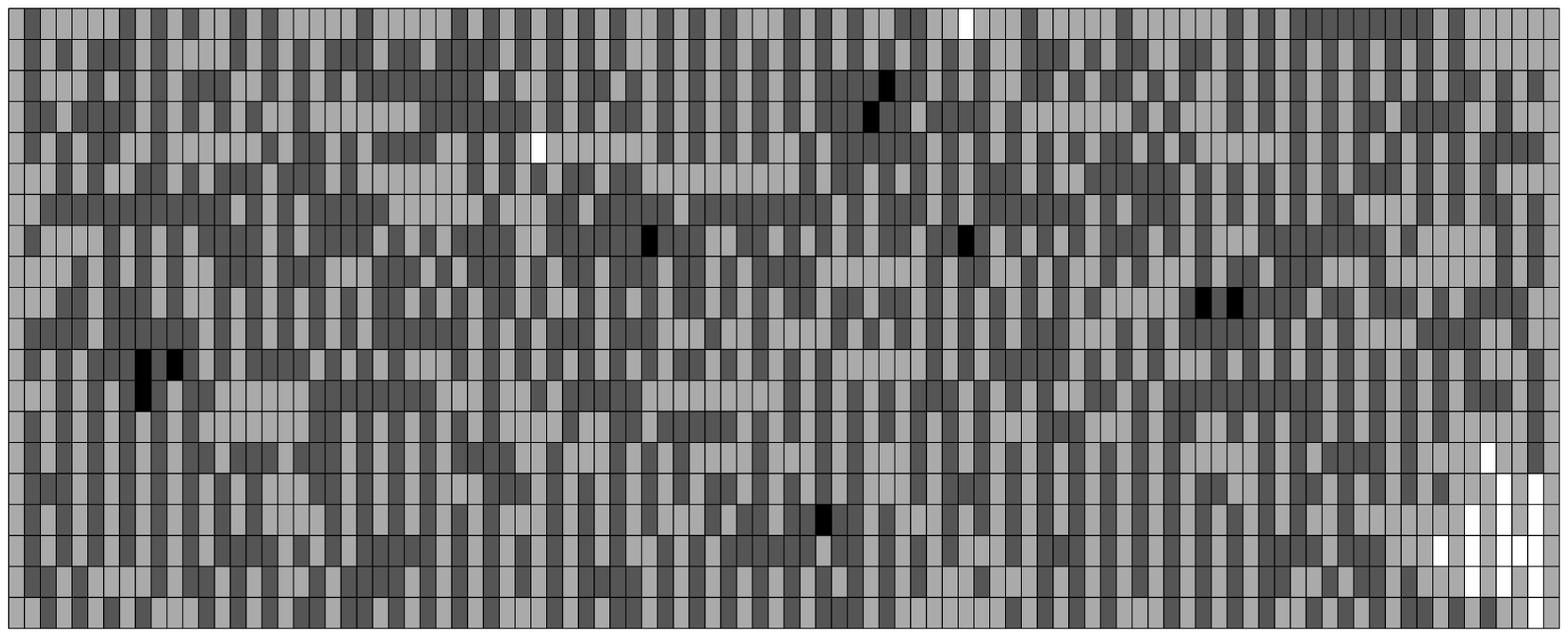,width=\linewidth}   
\end{figure}

%\vskip 1cm
\begin{figure}
\begin{picture}(20,60)
\end{picture}
\end{figure}
(b)
\vskip -4.5cm
\begin{figure}
\hspace{-2ex}\epsfig{figure=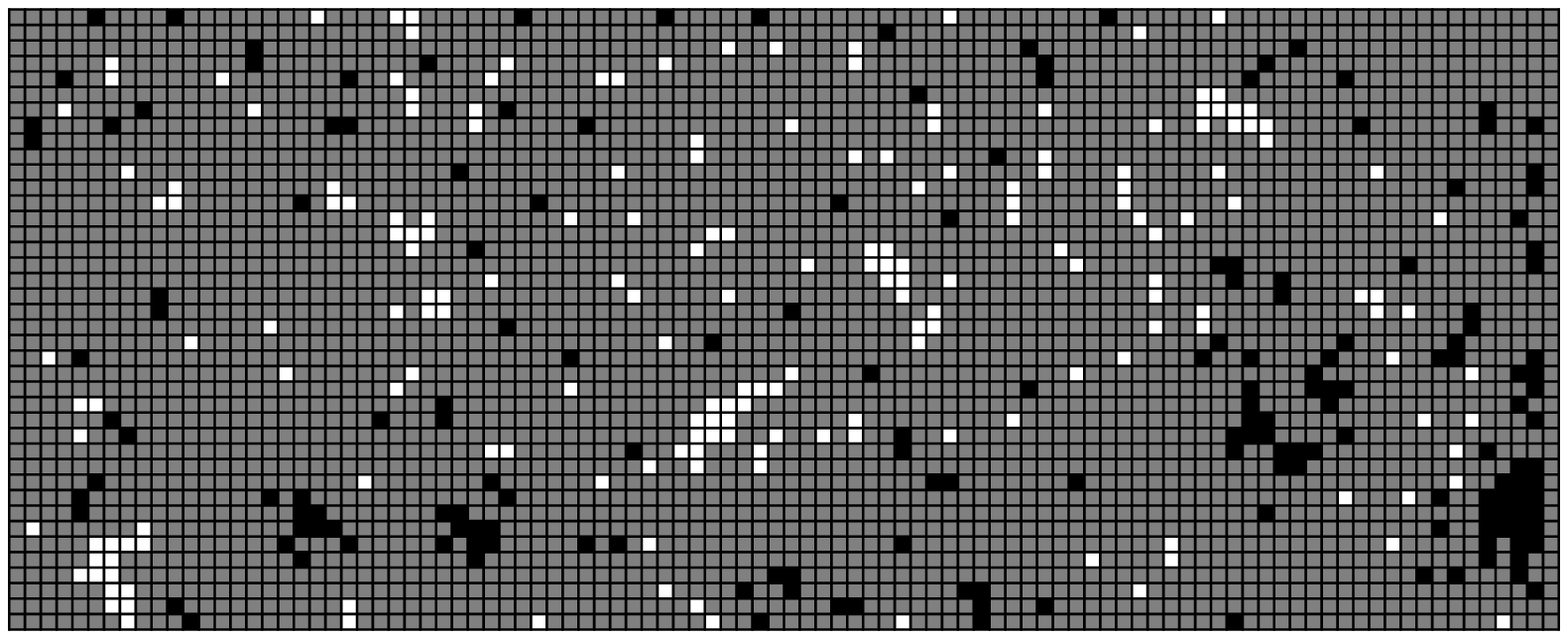,width=\linewidth}      
\end{figure}

\vskip -4.5cm
(c)
\vskip -4.5cm
\begin{figure}

\hspace{-3ex}\epsfig{figure=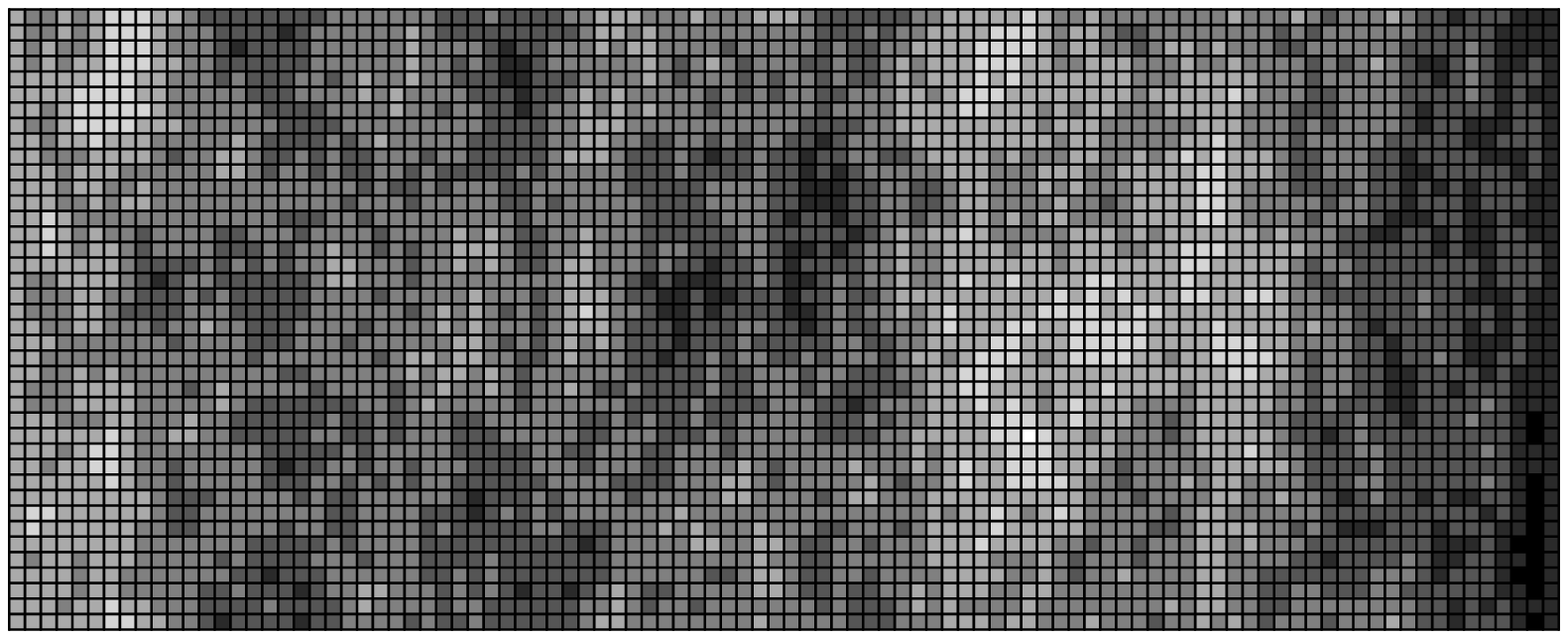,width=\linewidth}   
\end{figure}

\vskip -4.5cm
(d)
\vskip -0.3cm
\begin{figure}
%\vspace{0.5ex}
%\epsffile{dw.2.0.0.g.ps}      
%\epsffile{dw.05.-025.-1.g.ps}   
%\epsffile{dw.-1.05.2.g.ps}   
\hspace{-3ex}\epsfig{figure=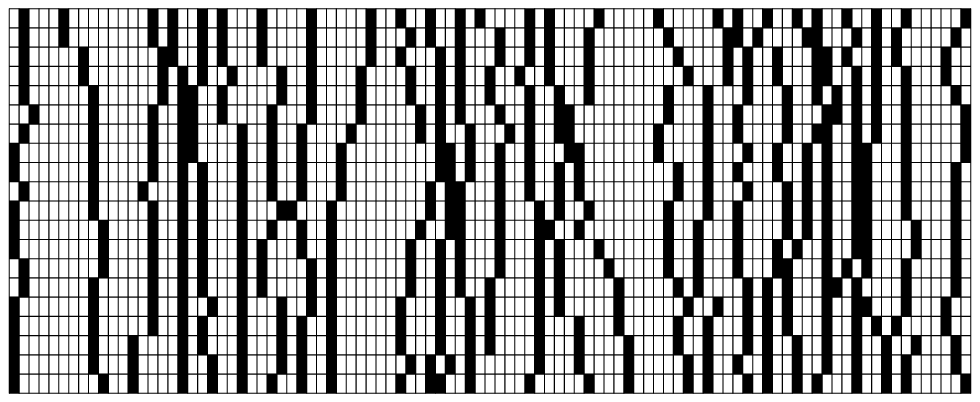,width=\linewidth}   
%\vspace{1.5ex}
\caption{ Monte-Carlo snapshots of (a) the Haldane phase (Phase V, 
$D=0,J=1,E=0$), (b)  the flat phase (Phase II, $D=2, J=0, E= 0$),
(c) the rough phase (Phase IV, $D=0, J=-0.5, E= 0$) and (d) the slanted phase
(Phase VII, $D=-0.75,J=-5,E=5$). Black to white means increasing height,
except for the slanted phase in (d), where black denotes a horizontal
link and white an up diagonal link.  }
\label{fig_snap}
\end{figure}

The fascinating order parameter of the Haldane phase or disordered
flat phase becomes transparent when looking at the worldsheet, using
the height representation instead of the spin-1 language.  Globally
the surface is limited to two heights only, and is therefore
macroscopically flat.  However in both the time and space direction
there is a disordered array of up and down steps, with the restriction
that every step up is followed by a step down, when the order is
perfect.  On flat pieces, however, there are local fluctuations (with
consecutive up steps or down steps) decreasing the value of the order
parameter.  These islands will grow when $J \rightarrow 0$ from the
positive side, and when they overlap the string becomes rough.  Note
that the rigidity (flatness) is clear when viewing the overall
structure of the worldsheet in 1+1D.  On the other hand, from a single
time slice one might be tempted to conclude that the string is rough.

The transition from phase
I to VII is of the Prokovsky-Talapov type.  Such transitions are
often discussed in the context of
commensurate-incommensurate transition of a monolayer of atoms on a
substrate with a different lattice parameter.  In the ``floating solid''
phase such a system consists of a set of parallel domain walls with entropic
meandering.\cite{dennijsrev} The similarity with phase VII,
illustrated in Fig.\ \ref{fig_snap} is clear.  The
entropic meandering in the 2D classical case is now to be interpreted as
the quantum motion of hard-core particles (horizontal links) along the
string.

\section{Discussion and conclusions}\label{sec_dis}

Motivated by stripes, 
we have introduced  a lattice string model for quantum domain
walls and mapped out its full phase diagram. We find a generic zero
temperature  
symmetry breaking: the string acquires a direction in all cases. 
The main reason is that bends
in the string prohibit the quantum transport, or, vice versa, the quantum
motion of kinks straightens out the string (the ``garden hose'' effect
of Nayak and Wilczek\cite{nawi}). We arrive at the
counter-intuitive conclusion that for increasing kink quantum disorder
the orientational preference of the string grows. The 
directed string problem which remains appears to be related to a well 
understood surface statistical physics (RSOS) model and simultaneously to a 
$S=1$ XXZ quantum spin chain with single site anisotropy. Motivated by the 
string interpretation, we found a number of phases described by this class 
of models which were previously not identified. 

Physically, the phases fall in three main categories: classical (flat
worldsheet), Gaussian (rough worldsheet) and `disordered flat' phases.
The phases are further distinguished by the direction they take in the
embedding space. Besides the flat strings in the horizontal and diagonal
directions, we find that the disordered flat phases show here a rich behavior. 
Apart from the known phase with horizontal direction,
which is associated with the incompressible phase of the spin model,
we identified a new category of disordered flat phases which take,
depending on
parameters, arbitrary directions in space (the ``slanted'' phases).

Although this does not apply to the localized strings, we suspect that a
strong universality principle might apply to the delocalized strings:
{\em At least away from the phase boundaries to the localized phases,
the underlying lattice renders the delocalized strings to be described
by free field theory.} The reason is simple: {\em regardless of the
terms that
one adds to the lattice scale action, the problem remains of the XXZ
kind and the massless phases fall into the 1+1D O(2) universality class.} 
For instance, one can add other kink-kink interactions, etcetera, and 
these can be all described by products of $S^z$ operators. Although these
operators determine the nature of the localized phases, they turn into
irrelevant operators in the massless phases. The kinetic sector is more
subtle. For instance, one would like to release the constraint that kinks
only occur with `height flavor' $\pm 1$. This means in surface language
that one partially lifts the restrictedness of the RSOS model, or in spin
language that one {\em increases the total spin}: e.g., $S=2$ means in 
string language that kinks occur describing height differences of
$\pm 2$ as well. Although increasing the magnitude of spin has an influence 
on the localized phases, it does not change the fact that the massless phase
away from the phase boundaries is still obeying XY universality. A point of
caution is that the `holes' in principle could change their order when larger
excursions are allowed. However, these `exchange loops' are strictly local
and therefore irrelevant for the long wavelength behavior as long as the
string is internally an insulator. These could represent more of a problem
for strings which are internally superconductors or metals.

We also stress that it follows from the arguments of den Nijs and
Rommelse\cite{nijsrom} that  the
occurrence of a gapped Haldane type phase for strings is not a
peculiar feature of the spin 1 representation, but a general
consequence of the existence of further neighbor interactions between
the holes in strings.
    
Do our findings bear any relevance to the stripes in cuprates? At the very 
least, they do bring up some interesting questions:\\
{\em (a)} Is the stripe solidification in for instance the LTT 
cuprates\cite{tranquada} in first instance driven by a single string effect 
or by a collective 
transition of the string liquid? In the end it has to be the latter, since
a single string cannot undergo phase transitions at finite temperatures.
However, it can be well imagined that the effect of the LTT-pinning potential
is to stabilize (1,0) directed stripes over (1,1) stripes. In the language 
of this paper, this amounts to an increase of the parameter $K$ which could
move the stripe from the Gaussian phase into the horizontal flat phase. At
zero temperature, this would turn individual stripes in straight rods which
are obviously much easier to order than meandering strings. At
finite temperatures, this could increase the single string persistence length 
substantially, so that stripe-stripe interactions become more effective 
in stabilizing a stripe solid at finite temperatures\cite{Cop82}.
 Further work is
needed to establish if these single stripe transitions are of relevance.\\
{\em (b)} Do the `disordered flat' string phases exist? The simplest disordered
flat phase is the horizontal one (phase V) corresponding with the Haldane
phase of the $S=1$ spin chain. In string language, this is nothing else
than a localized string along the (1,0) direction in the lattice which
is however not site-centered (as phase II) but, on average, {\em bond centered}. 
Bond centered stripes show up in the numerical study of the $t$-$J$ model by 
White and Scalapino\cite{whisca}, which 
shows that the ground state of this model at finite dopings is a stripe
phase. A main difference with the mean-field stripes is that these $t$-$J$
stripes are bond centered. In first instance, one could be tempted to
think that this has a truly microscopic reason: charges in $t$-$J$ prefer
to live on links. However, it could also be due to a {\em collective
string effect} --- it could be ``our'' phase V. This can be easily established 
by measuring the appropriate (string) correlators. Is it so that on equal
times the charges live on sites while the kinks take care of delocalizing
the stripes over two lattice rows, or is it so that on all times the charges
are living on the links? This is obviously an important question in the light
of recent works relating the bond centering via Hubbard-ladder physics to
superconductivity\cite{dhlee}. 
We also notice that there are experimental indications
for bond-centering in the nickelates\cite{nickel1}
 where disordered flatness could
possibly also play a role.\\
{\em (c)} If well developed stripes exist in the superconductors and/or metals,
these have to occur in the form of a quantum disordered stripe phase,
or a `quantum string liquid'. What is learnt in this regard from the 
present study of a single string? A prerequisite for the existence of
a quantum string liquid is that a single string is delocalized. If our
conjecture that a single critical string is described by free field theory
turns out to be correct,
this amounts to a considerable simplification. In Euclidean space time,
the single free string worldsheet is like a Gaussian membrane and a
system of strings becomes a system of interacting Gaussian membranes,
embedded in 2+1 dimensions. This in turn is like a classical incommensurate
system in 3D which, although barely studied, appears as a tractable
problem. For instance, it is known that the 3D incommensurate solid
melts at a finite temperature in all cases\cite{fischer2}.
For the quantum case this 
means that the quantum-melting transition will occur at some finite
value of the coupling constant, which in turn depends on the single
string quantum fluctuation as well as string-string interaction effects.
Investigations addressing this many string problem are in progress, profiting 
from the simple fluctuation behavior of a single string.

\acknowledgments
 We thank S. A. Kivelson  for 
stimulating discussions.
The work of HE was supported by the Stichting voor Fundamenteel Onderzoek der Materie
(FOM), which is financially supported by the Nederlandse Organisatie voor
Wetenschappelijk Onderzoek (NWO), and JZ acknowledges support by the Dutch
Royal Academy of Sciences (KNAW).

\begin{figure}
\begin{picture}(20,60)
\end{picture}
\end{figure}

%\appendix
%\vspace*{4cm}

{\bf APPENDIX.}

\vspace*{0.5cm}

In this appendix we give the $t$-matrix elements of the blocks
defined in Fig.\ \ref{fig_Tblocks}. The local $t$-matrix
at position $l$ and Trotter slice $k$ is defined as,
\begin{equation}
    \langle  {\bf r}_{l-1,k}, {\bf r}_{l,k}, {\bf r}_{l+1,k} |
    \exp{ \frac{1}{n}  {\cal H}_l }
    | {\bf r}^\prime_{l-1,k}, {\bf r}^\prime_{l,k}, {\bf r}^\prime_{l+1,k},
    \rangle.
        \label{app_tmat}
\end{equation}
The matrix elements depend on the positions of three members of the
string, $l-1$, $l$ and $l+1$.  The positions of $l-1$ and $l+1$ are
required to be identical ${\bf r}_{l-1,k} = {\bf r}^\prime_{l-1,k}{\bf
r}^\prime_{l-1,k}$ in the two Trotter subslices involved, due to the
checkerboard decomposition, but the position of member $l$ can be
different, leading to off-diagonal matrix elements.  The matrix
elements of the $t$-matrix are easily found by first diagonalizing
${\cal H_l}$ and expanding the basis vectors in terms of the
eigenvectors.

Block A: This block contains three configurations, see Fig.\
\ref{fig_Tblocks}.  We use the same order for the states as in the
figure.  Note that only half of the energy ${\cal K}$ of the diagonal
link between $l-1$ and $l$ and the link between $l$ and $l+1$ should
be contributed to $l$.  The Hamiltonian,
\begin{equation}
        {\cal H} = \left(
        \begin{tabular}{ccc}
                ${\cal K}$ & ${\cal T}$ &      0      \\
                ${\cal T}$ &      0     & ${\cal T}$  \\
                     0     & ${\cal T}$ & ${\cal K}$  \\
        \end{tabular}
        \right),
        \label{app_A_H}
\end{equation}
is easily diagonalized. The eigenvalues are ${\cal K}$, $E_+$ and
$E_-$. The $t$-matrix is,
\begin{eqnarray}
    t &=& \left(
        \begin{tabular}{ccc}
                $t_{11}$  &  $t_{12}$  &  $t_{13}$  \\
                $t_{12}$  &  $t_{22}$  &  $t_{12}$  \\
                $t_{13}$  &  $t_{12}$  &  $T_{11}$  \\
        \end{tabular}
        \right) ,                                         \nonumber \\
    t_{11} &=& {1 \over 2} e^{{\cal K}/n} + N^2_+ e^{E_+/n}
                                       + N^2_- e^{E_-/n}, \nonumber \\
    t_{12} &=&  N^2_+ \alpha_+ e^{E_+/n}
            + N^2_- \alpha_- e^{E_-/n} ,                  \nonumber \\
    t_{13} &=& -\frac{1}{2} e^{{\cal K}/n} + N^2_+ e^{E_+/n}
                                        + N^2_- e^{E_-/n},\nonumber \\
    t_{22} &=&  N^2_+ \alpha^2_+ e^{E_+/n}
                                       + N^2_- e^{E_-/n}, \nonumber \\
    E_{\pm} &=& \frac{}{2} \pm
              \frac{\sqrt{{\cal K}^2 + 8 {\cal T}^2}}{2}, \nonumber \\
    \alpha_{\pm} &=& \frac{ E_{\pm} - {\cal K} }{{\cal T}},
    \hspace{2em}  N_{\pm} = \frac{1}{\sqrt{2+\alpha^2_{\pm}}}.
        \label{app_A_t}
\end{eqnarray}
Here $n$ is the number of Trotter slices and ${\cal K}$,
${\cal L}_{12}$, ${\cal L}_{22}$, ${\cal L}_{11}$ and ${\cal T}$
are the string model parameters.

Block B contains two configurations, each with one horizontal and one
diagonal link.  Repeating the above procedure one finds,
\begin{eqnarray}
    t &=& \left(
        \begin{tabular}{cc}
        $e^D \cosh({\cal T}/n)$  &  \hspace{1em}$e^D \sinh({\cal T}/n)$   \\
        $e^D \sinh({\cal T}/n)$  &  \hspace{1em}$e^D \cosh({\cal T}/n)$   \\
        \end{tabular}
        \right) ,                                         \nonumber \\
    D &=& { {\cal K} \over 2 n } + {{\cal L}_{12} \over n}
        \label{app_B_t}
\end{eqnarray}

Block C contains a single configuration of two diagonal links, and
the energy and $t$-matrix therefore contain ${\cal L}_{22}$,
\begin{equation}
    t = \exp( { {\cal K} \over n } + {{\cal L}_{22} \over n} ) ,
        \label{app_C_t}
\end{equation}

Block D consists of a square corner between one horizontal and one
vertical link, and ${\cal L}_{11}$ is involved,
\begin{equation}
    t = \exp( {{\cal L}_{11} \over n} ) ,
        \label{app_D_t}
\end{equation}

%\narrowtext
\begin{figure}
\begin{picture}(20,160)
\end{picture}
\end{figure}

\begin{table}[tbp]
\hspace{0.5cm} \caption{ Order parameters that distinguish between the
  six different phases in the phase diagram for $E=0$. A + entry in
  the table indicates that the particular order parameter is nonzero.}
%\hspace{0.5cm}
%\vspace{-0.2cm}
\begin{tabular}[t]{ccccc}
\hline
\hline
  Phase   & \hspace{0.7cm} $\rho$  & \hspace{0.7cm} $\rho_{stag}$ & \hspace{0.7cm} 
$\rho_{str}$ & \hspace{0.7cm} $\rho_{slope}$
  \\
\hline
  I   &   & \hspace{0.5cm} +  &    & \hspace{0.5cm}  +  \\
  II  & \hspace{0.5cm} + &    &    &   \\
III   &   & \hspace{0.5cm} +  & \hspace{0.5cm} +   &   \\ 
IV    &   &    &    &   \\
V     &   &    & \hspace{0.5cm}  + &  \\
VI    &   & \hspace{0.5cm} +  &    &  \\
\hline
\hline
\end{tabular}
\protect\label{tabE0Phases}
\end{table}
 
\begin{table}[tbp]
\caption{Spin 1 $S$ seen as a combination of two spins 1/2, $\sigma$
and $s$.}
\begin{tabular}{cccc}
\hline
\hline
        $S$      & \hspace{2cm}     1     &  \hspace{2cm}      0      & 
 \hspace{2cm}    -1        \\
        \hline
        $\sigma$ &  \hspace{2cm} $\uparrow$ &  \hspace{2cm} $\downarrow$ & 
 \hspace{2cm} $\uparrow$    \\
        $s$      &  \hspace{2cm} $\uparrow$ &  \hspace{2cm}      -    &  
 \hspace{2cm} $\downarrow$  \\
\hline
\hline
\end{tabular}
\protect\label{tabSpinhalf}
\end{table}

\setlength{\unitlength}{0.1cm}
\begin{table}
\hspace{0.5cm} \caption{ A schematic representation of the different
phases.  Also shown is the long-range order of the two spins 1/2, $s$
and $\sigma$ as defined in the text.  F = ferromagnetic, F1 = up-spin
ferromagnetic, F2 = down-spin ferromagnetic, AF = antiferromagnetic, D
= disordered }
%\hspace{0.5cm}
%\vspace{-0.2cm}
\begin{tabular}[t]{ccccc}
\hline
\hline
  Phase   & \hspace{0.4cm}  $\sigma$  & \hspace{0.4cm}  s  &   \hspace{0.4cm} 
String  &  \hspace{0.4cm}  Spin 1   \\
\hline
    I     & \hspace{0.4cm}  F1     &  \hspace{0.4cm}   F    &  \hspace{0.4cm}
 \begin{picture}(10,10)
 \multiput(0,0)(2,2){4}{\circle*{0.9}}
\end{picture}
                              &  \hspace{0.4cm}  ++++++++   \\
       II    &  \hspace{0.4cm}   F2   & \hspace{0.4cm}   -  &  \hspace{0.4cm}
\begin{picture}(18,4)
 \multiput(0,0)(2,0){9}{\circle*{0.9}}
\end{picture}
                              &  \hspace{0.4cm}  0 0 0 0 0 0 0 0    \\
       III   &  \hspace{0.4cm}   F1   & \hspace{0.4cm} AF &  \hspace{0.4cm}
\begin{picture}(18,6)
 \multiput(0,0)(4,0){5}{\circle*{0.9}}
 \multiput(2,2)(4,0){4}{\circle*{0.9}}
\end{picture}
                              &  \hspace{0.4cm}  +$-$+$-$+$-$+$-$     \\
       IV    &  \hspace{0.4cm}   D   &  \hspace{0.4cm}  D  & \hspace{0.4cm}
\begin{picture}(20,8)
 \put(0,0){\circle*{0.9}}
 \put(2,2){\circle*{0.9}}
 \put(4,2){\circle*{0.9}}
 \put(6,0){\circle*{0.9}}
 \put(8,2){\circle*{0.9}}
 \put(10,2){\circle*{0.9}}
 \put(12,4){\circle*{0.9}}
 \put(14,4){\circle*{0.9}}
 \put(16,2){\circle*{0.9}}
 \put(18,4){\circle*{0.9}}
\end{picture}
                              & \hspace{0.4cm}  +0$-$+0+0$-$+   \\
       V      & \hspace{0.4cm}    D   &  \hspace{0.4cm}  AF & \hspace{0.4cm}
\begin{picture}(16,6)
 \put(0,2){\circle*{0.9}}
 \put(2,0){\circle*{0.9}}
 \put(4,2){\circle*{0.9}}
 \put(6,2){\circle*{0.9}}
 \put(8,0){\circle*{0.9}}
 \put(10,0){\circle*{0.9}}
 \put(12,0){\circle*{0.9}}
 \put(14,2){\circle*{0.9}}
 \put(16,0){\circle*{0.9}}
\end{picture}
                              & \hspace{0.4cm}  $-$+0$-$0 0+$-$     \\
       VI      &  \hspace{0.4cm}   F1   &  \hspace{0.4cm}  D  & \hspace{0.4cm}
\begin{picture}(16,8)
 \put(0,0){\circle*{0.9}}
 \put(2,2){\circle*{0.9}}
 \put(4,0){\circle*{0.9}}
 \put(6,2){\circle*{0.9}}
 \put(8,4){\circle*{0.9}}
 \put(10,2){\circle*{0.9}}
 \put(12,4){\circle*{0.9}}
 \put(14,2){\circle*{0.9}}
 \put(16,0){\circle*{0.9}}
\end{picture}
                              & \hspace{0.4cm}  +$-$++$-$+$-$$-$     \\
       VII     &  \hspace{0.4cm}   D   &  \hspace{0.4cm}  F  & \hspace{0.4cm}
\begin{picture}(10,10)
 \put(0,0){\circle*{0.9}}
 \put(2,2){\circle*{0.9}}
 \put(4,2){\circle*{0.9}}
 \put(6,4){\circle*{0.9}}
 \put(8,6){\circle*{0.9}}
 \put(10,6){\circle*{0.9}}
\end{picture}
                              & \hspace{0.4cm} 0+0++0+0 0     \\
       VIII     &  \hspace{0.4cm}   AF  &  \hspace{0.4cm}  F & \hspace{0.4cm}
\begin{picture}(14,10)
 \put(0,0){\circle*{0.9}}
 \put(2,0){\circle*{0.9}}
 \put(4,2){\circle*{0.9}}
 \put(6,2){\circle*{0.9}}
 \put(8,4){\circle*{0.9}}
 \put(10,4){\circle*{0.9}}
 \put(12,6){\circle*{0.9}}
 \put(14,6){\circle*{0.9}}
\end{picture}
                              & \hspace{0.4cm}  0+0+0+0+0     \\
       IX       &  \hspace{0.4cm}   AF  &  \hspace{0.4cm}  D  & \hspace{0.4cm}
\begin{picture}(18,8)
 \put(0,0){\circle*{0.9}}
 \put(2,0){\circle*{0.9}}
 \put(4,2){\circle*{0.9}}
 \put(6,2){\circle*{0.9}}
 \put(8,4){\circle*{0.9}}
 \put(10,4){\circle*{0.9}}
 \put(12,2){\circle*{0.9}}
 \put(14,2){\circle*{0.9}}
 \put(16,4){\circle*{0.9}}
 \put(18,4){\circle*{0.9}}
\end{picture}
                              & \hspace{0.4cm}  0+0+0$-$0+0     \\
       X      &   \hspace{0.4cm}  AF  &  \hspace{0.4cm}  AF  & \hspace{0.4cm}
\begin{picture}(14,6)
 \put(0,0){\circle*{0.9}}
 \put(2,0){\circle*{0.9}}
 \put(4,2){\circle*{0.9}}
 \put(6,2){\circle*{0.9}}
 \put(8,0){\circle*{0.9}}
 \put(10,0){\circle*{0.9}}
 \put(12,2){\circle*{0.9}}
 \put(14,2){\circle*{0.9}}
 \put(16,0){\circle*{0.9}}
 \put(18,0){\circle*{0.9}}
\end{picture}
                              & \hspace{0.4cm}  0+0$-$0+0$-$0     \\

\hline
\hline
\end{tabular}
\protect\label{tabPhases}
\end{table}

\end{multicols}


\begin{references}
\bibitem{selfavoid} In the high temperature limit, the string becomes
  a self-avoiding walk on a two-dimensional lattice. For such a walk
  of length $N$,
  the radius of gyration $R_g$ grows as $R_g \sim N^{\nu}$ where
  $\nu=0.6 $ in D=2. This implies that the directedness defined in
  this paper should go to zero as $N^{\nu-1}= N^{-0.4}$ for
  $N\rightarrow \infty$ in the high
  temperature limit .
\bibitem{noteheight} These order parameters, as well as the string
  order parameter introduced below, are only nonzero once the height
  at an arbitrary point is specified. This can, e.g., be the guider
  point discussed in section V or the reference height $h_0$
  introduced in Ref. \onlinecite{nijsrom}.
\end{references}
\end{document}